\documentclass[aps,pre,twocolumn,superscriptaddress,longbibliography]{revtex4-1}

\usepackage{graphicx}
\usepackage{epsfig}
\usepackage{dcolumn}
\usepackage{bm}
\usepackage{mathrsfs}
\usepackage{multirow}
\usepackage[all]{xy}
\usepackage{pbox}
\usepackage{amssymb}

\usepackage{color}
\definecolor{myblue}{RGB}{65,105,225}
\definecolor{mygreen}{RGB}{34,139,34}
\definecolor{myorange}{RGB}{255,69,0}
\definecolor{mygray}{RGB}{211,211,211}
\definecolor{mywhite}{RGB}{255,255,255}
\usepackage[colorlinks,linkcolor=myorange,urlcolor=myblue,citecolor=myblue]{hyperref}

\usepackage{multirow, makecell}

\def\(({\left(}
\def\)){\right)}
\def\[[{\left[}
\def\]]{\right]}

\newcommand{\beq}{\begin{equation}}
\newcommand{\eeq}{\end{equation}}

\newcommand{\ben}{\begin{eqnarray}}
\newcommand{\een}{\end{eqnarray}}

\newcommand{\angstrom}{\mbox{\normalfont\AA}}

\newcommand{\la}{\langle}
\newcommand{\ra}{\rangle}
\newcommand{\be}{\begin{equation}}
\newcommand{\ee}{\end{equation}}

\let\vec\mathbf

\begin{document}  

\title{Molecular hints of two-step transition to convective flow via streamline percolation}

\author{P.L. Garrido}
\email[]{garrido@onsager.ugr.es}
\affiliation{Departamento de Electromagnetismo y F\'{\i}sica de la Materia, and Institute Carlos I for Theoretical and Computational Physics, Universidad de Granada, Granada 18071, Spain}

\author{P.I. Hurtado}
\email[]{phurtado@onsager.ugr.es}
\affiliation{Departamento de Electromagnetismo y F\'{\i}sica de la Materia, and Institute Carlos I for Theoretical and Computational Physics, Universidad de Granada, Granada 18071, Spain}

\date{\today}

\begin{abstract}
Convection is a key transport phenomenon important in many different areas, from hydrodynamics and ocean circulation to planetary atmospheres or stellar physics. However its microscopic understanding still remains challenging. Here we numerically investigate the onset of convective flow in a compressible (non-Oberbeck-Boussinesq) hard disk fluid under a temperature gradient in a gravitational field. We uncover a surprising two-step transition scenario with two different critical temperatures. When the bottom plate temperature reaches a first threshold, convection kicks in (as shown by a structured velocity field) but gravity results in hindered heat transport as compared to the gravity-free case. It is at a second (higher) temperature that a percolation transition of advection zones connecting the hot and cold plates triggers efficient convective heat transport. Interestingly, this novel picture for the convection instability opens the door to unknown piecewise-continuous solutions to the Navier-Stokes equations. 
\end{abstract}

\maketitle 

\section{Introduction}

When a fluid is heated from below in a gravitational field, there exists a critical temperature beyond which it develops a distinctive roll flow pattern which transports cold fluid from the upper, denser layers to the hotter, lighter regions near the bottom plate and viceversa. This is the well-know Rayleigh-B\'enard (RB) convection phenomenon \cite{bodenschatz00a,ahlers09a}, one of the simplest and most studied instabilities in fluid dynamics \cite{mutabazi10a}. The importance of convection is difficult to overstate. It plays a key role in a wide range of phenomena, from e.g. atmospheric \cite{ogura62a,gierasch00a} and oceanic \cite{marshall99a} circulation or planetary mantle dynamics \cite{morgan71a,davies92a} to stellar physics \cite{roxburgh78a,meakin07a,schumacher20a} or granular flow \cite{murdoch13a,pontuale16a}, to mention just a few. Convection can also be harnessed for technological applications such as e.g. cooling solutions in nanoelectronics \cite{bar-cohen07a,iyengar14a} and heat sinks \cite{yu10a,kalteh12a,teertstra00a}, convection ovens \cite{verboven00a} or metal-production processes \cite{calmidi00a,brent88a}. At the fundamental level, convection has been instrumental in the development of stability theory in hydrodynamics \cite{chandrasekhar13a}, and a paradigm for pattern formation \cite{getling98a,bodenschatz00a} and spatiotemporal chaos \cite{ahlers09a}. Moreover, recent advances in convection physics include large-scale circulation \cite{brown05a,brown07a} and superstructures \cite{pandey18a,hartlep03a} in convective flow, steady-state degeneracy \cite{wang20a} or the role of periodic modulations \cite{yang20a}. However, and despite more than a century of detailed investigation, the microscopic understanding of this ubiquitous fluid instability still remains elusive. 

At the theoretical level the RB instability can be rationalized within the Oberbeck-Boussinesq (OB) approximation to the Navier-Stokes equations \cite{landau13a,chandrasekhar13a,gray76a}, which assumes that the properties of the fluid do not depend on the local temperature, except for the density profile (for which a linear approximation is used). This approximation has also been used to understand the role of fluctuations on the convection transition \cite{zaitsev71a,graham74a,swift77a,wu95a}, becoming the standard theoretical framework to understand the RB instability. 
Indeed, in some contexts  the term \emph{RB convection} is meant to encompass both the RB set up and the OB theoretical approximation. Here however we prefer to distinguish the physical phenomenon from the theoretical framework traditionally used to describe it. In fact, the range of validity of the strong assumptions underlying Oberbeck-Boussinesq theory is still unclear, as demonstrated e.g. by the appearance of non-Oberbeck-Boussinesq effects \cite{ahlers09a,pandey21a} and deviations for compressible fluids \cite{bormann01a,risso93a}, thus demanding an atomistic, microscopic assessment of the instability. Early heroic works in molecular dynamics studied the RB instability from a microscopic point of view \cite{mareschal87a,mareschal87b,mareschal88a,rapaport88a,puhl89a,rapaport92a,risso93a}, though large velocity fluctuations prevented the accurate measurement of local observables and the precise characterization of the transition. Modern computer power and techniques have overcome these issues, see e.g. \cite{rapaport06a,pandey18a,yang20a,yang20b}, but there is a surprising lack of detailed studies of the onset of convection from a microscopic viewpoint. This work fills this gap by numerically investigating the nature of the convection instability in a two-dimensional compressible (non-Oberbeck-Boussinesq) hard disk fluid under a temperature gradient in a gravitational field. Strikingly, we find a clear-cut two-step transition scenario with two different critical temperatures. Initially, convection kicks in when the bottom plate temperature reaches a first threshold, as demonstrated by a structured but still roll-free velocity field. At this point local coherent motions emerge, but they are disordered and disconnected and hence unable to promote energy transfer against the gravitational field. We call this regime \emph{semi-convective} due to its inefficient transport properties as compared to the gravity-free case. Increasing the bottom plate temperature, a second critical point is reached where efficient heat transport is triggered, leading to a \emph{fully-convective regime}. We show that this second critical point, at which a clear roll structure emerges, is related to an underlying percolation transition of advection zones connecting the hot and cold plates. Note that related percolation phenomena have been recently linked to different transitions to turbulent behavior, as e.g.~in Couette and pipe flows  \cite{avila11a,shi13a,lemoult16a,klotz22a}. Interestingly, the emergence of structured but disordered flow patterns in the semi-convective regime, also hinted at in experiments \cite{wu95a} and simulations \cite{zhang09a,zhang17b} on fluctuations below the RB instability, suggests the existence of unknown piecewise-continuous solutions to the Navier-Stokes equations in between the two critical temperatures. Finally, we stress that the existence of a semi-convective, gravity-suppressed transport regime seems to be linked to the compressibility of the underlying flow field.

\section{Model and simulation}

Hard particle systems are among the most inspiring, successful and prolific models of physics, as they contain the key ingredients to understand a large class of complex phenomena \cite{mulero08a,szasz00a,hurtado20a}. Here we consider a two-dimensional fluid of $N$ hard disks moving in a square box of unit side $L=1$ under the action of a constant gravitational field $\vec g=(0,-g)$. Each disk has unit mass ($m=1$) and a radius $\tilde{r}$ chosen so that the packing fraction $\rho=N\pi \tilde{r}^2/L^2$ is fixed, i.e. $\tilde{r}=[\rho L^2/(N\pi)]^{1/2}$. Disks move freely in between collisions along trajectories $\vec r_i(t_k+\Delta t)=\vec r_i(t_k)+\vec v_i(t_k)t+\frac{1}{2}\vec g \Delta t^2$ and $\vec v_i(t_k+\Delta t)=\vec v_i(t_k)+\vec g \Delta t$, where $\vec r_i(t_k)$ and $\vec v_i(t_k)$ are the values of the $i^{\text{th}}$-disk position and velocity, respectively, \emph{right after} its last collision event at time $t_k$. Disks collisions are elastic, conserving both energy and linear momentum, and no internal rotational degrees of freedom are considered. In addition, boundary conditions in the square box include reflecting walls along the $x$-direction and stochastic thermal walls \cite{livi17a,dhar08a,lepri03a,bonetto00a} at the top and bottom boundaries at temperatures $T_0$ and $T$, respectively. Note that for any non-zero temperature gradient $\Delta T\equiv|T-T_0|/L$ we expect a net heat current flowing from the hot plate to the cold one \cite{livi17a,dhar08a,lepri03a,bonetto00a}.

The focus of this work is not on hydrodynamic pattern formation \cite{rapaport06a} but instead on the accurate, detailed measurement of different order parameters and transport properties across the RB instability. We hence choose a moderate number of particles, $N=957$, in order to gather the extensive amounts of data needed to perform significant averages. Hard-sphere systems of $N={\cal O}(10^3)$ particles have been shown to exhibit clear macroscopic hydrodynamic behavior fully compatible with (nonlinear) Navier-Stokes equations \cite{mareschal87a,mareschal87b,mareschal88a,rapaport88a,puhl89a,rapaport92a,risso93a,pozo15b,pozo15a,hurtado16a,hurtado20a}. Indeed, the relevance of our results in the large system size limit is strongly supported by the recently discovered bulk-boundary decoupling phenomenon in hard particle systems \cite{pozo15b,pozo15a,hurtado16a}, which enforces the macroscopic hydrodynamic laws in the bulk of the computational finite-size fluid. Nevertheless, the absence of noticeable finite-size effects has been carefully checked by measuring in equilibrium ($T_0=T$) both hydrostatic density and pressure profiles and the local equation of state for different $g$'s (see Fig. \ref{figEq} in Appendix \ref{appA2} and related discussion), all exhibiting excellent agreement with macroscopic formulae. 

Initially the disks are placed regularly on the box with an initial velocity vector of random orientation and modulus $\sqrt{T_0+T}$. We then evolve the system during $5\times 10^4$ collisions per particle (we set the time unit to one collision per particle on average) to guarantee that the correct steady state has been reached. Only then measurements of the different local and global observables start, every $100$ collisions per particle, for a total time of $10^7$ collisions per particle ($\sim 10^{10}$ total collisions), thus collecting a total of $M\sim 10^5$ data for averages. We use a $3\sigma$-convention for errorbars, so observables have typical errors of $3\sigma/\sqrt{M}\simeq 0.01 \sigma$ that suffice to accurately analyze the emergent behavior. For local measurements, we divide the unit box ($L=1$) into $n_c\times n_c$ virtual square cells of side $\Delta=1/n_c$, each one labeled by its center position $\vec{r}$, and we use here $n_c=30$ to capture the fine details of the hydrodynamic fields. The resulting total number of cells is comparable to $N$, so on average a cell contains only few particles in each snapshot of the dynamics, thus requiring a very large number of measurements to distill the relevant hydrodynamic behavior. Note however that, despite being far from the \emph{ideal} continuum limit of fluid dynamics \cite{landau13a}, the excellent self-averaging properties of the hard disk fluid \cite{pozo15b,pozo15a,mulero08a,szasz00a,mulero08a,hurtado20a} guarantee the proper convergence of local averages to the macroscopic behavior predicted by nonlinear hydrodynamics. In particular, we measure the average hydrodynamic velocity field $\la \vec u(\vec r)\ra$ and the packing fraction field $\la \eta(\vec r)\ra$ among other magnitudes, as well as global hydrodynamic observables as the total hydrodynamic kinetic energy $\la e\ra = \frac{\Delta^2}{2\rho}\sum_{\vec{r}} \la\eta(\vec{r})\ra \la \vec u(\vec r)\ra ^2$ or the heat current $\la J_g\ra$ traversing the fluid, which can be measured as the kinetic energy exchange in either thermal wall. In addition, we also measure some molecular observables as the average kinetic energy per particle in the fluid, $\la \varepsilon\ra = N^{-1} \sum_{i=1}^N\frac{1}{2} \la \vec{v}_i^2\ra$, as well as its variance $\sigma^2(\varepsilon)$. Appendix \ref{appA} describes some details on the measurement of these and other local and global observables. 

To choose the range for $(\rho,g,T_0,T)$ where the phenomenology of interest may emerge, we explore parameter space using linearized Navier-Stokes equations under the Oberbeck-Boussinesq approximation \cite{landau13a,chandrasekhar13a} together with the Henderson equation of state \cite{henderson77a} and the Enskog transport coefficients  \cite{mulero08a,hurtado20a} for hard disks, see Appendix \ref{appB}. This exploration suggests performing measurements for a fixed packing fraction $\rho=0.2$, gravity fields $g=5$, 10 and 15 (together with the gravity-free case $g=0$), 
and temperatures $T_0=1$ and $T=1,1.2,1.4,1.6,1.8, 2.0, 2.2, 2.4, 2.6, 2.8, 3, 4, \ldots, 19, 20$ for the top and bottom plates, respectively. 

To put our simulations into context, note that if we assume a physical disk radius of ${\tilde r}=3 \angstrom$, then the corresponding simulation box lengthscale is $L=367.8 \angstrom$ for a packing fraction $\rho=0.2$ and $N=957$ disks. Moreover, the maximum relative temperature gradient $(T-T_0)/T_0=19$ in our simulations corresponds to a physical temperature gradient of order $\Delta T\equiv|T-T_0|/L\approx 1.55 \times 10^{11}$ K/m assuming room temperature conditions $T_0\approx 300$ K and the previous physical value for $L=367.8 \angstrom$. These extreme conditions differ quantitatively from those observed in typical convective fluids, but still lead to flow fields that resemble those of real fluids \cite{rapaport88a}, being at the same time the optimal conditions where to explore in fine detail the convection instability from a fundamental, microscopic viewpoint. 

Note also that under such strong driving the hard disk fluid is compressible and the resulting hydrodynamic fields are highly nonlinear, with a nontrivial, asymmetric structure in the gradient direction (see Figs.~\ref{figs75}-\ref{figrho} below). Moreover the nonlinear dependence of the different transport coefficients on the local hydrodynamic fields becomes apparent and essential to understand the emergent macroscopic behavior \cite{pozo15b,pozo15a}. These effects signal a clear departure from the standard Oberbeck-Boussinesq hydrodynamic theory of the convection instability, making difficult the estimation of the different dimensionless numbers (as e.g. Nusselt or Knudsen numbers, etc.) which are typically used to characterize the instability within this approximation. However, in order to facilitate the connection with the classical hydrodynamic description, we can define \emph{locally} some of these adimensional numbers \cite{pandey21a} (as e.g. the Nusselt or Knudsen numbers), though the nonlinear effects described above will lead to depth variations in these numbers. We hence provide in Appendix \ref{appB2} local estimations of both the Knudsen and Nusselt numbers near the center of the simulation box for three representative state points of the fluid, and discuss their values in light of our results. We also report the Nusselt number near the top plate for different state points to discuss its depth variations.

\begin{figure}
\includegraphics[width=8.5cm,clip]{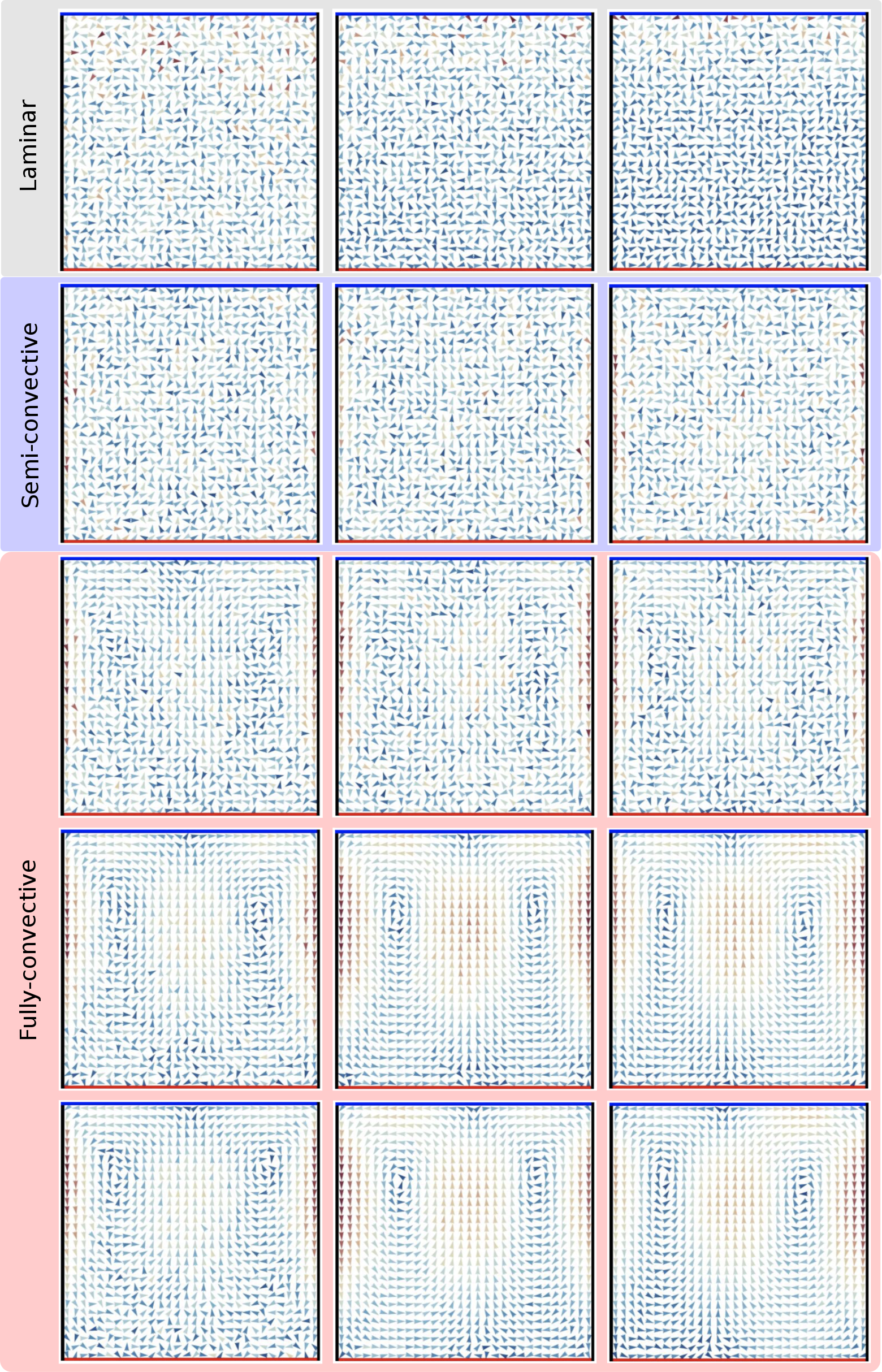}
\caption{{\bf Onset of convection.} Average hydrodynamic velocity field $\la \vec u(\vec r)\ra$ for hard disks measured across the Rayleigh-B\'enard instability. Color indicates velocity modulus, from $u_{\text{min}}$ (blue) to $u_{\text{max}}$ (red). The $(u_{\text{min}},u_{\text{max}})$ interval changes among panels. Left column: $g=5$ and $T=1.4, 2.6, 4,10, 20$, for which $(u_{\text{min}},u_{\text{max}})=$(0.000043, 0.016), (0.00025, 0.025), (0.00021, 0.04), (0.0016, 0.11), (0.0032, 0.15). Central column: $g=10$ and $T=1.8, 4, 6, 15, 20$, for which $(u_{\text{min}},u_{\text{max}})=$(0.00018, 0.029), (0.00017, 0.031), (0.001, 0.045), (0.0087, 0.24), (0.004, 0.3). Right column: $g=15$ and $T=2, 6, 8, 15, 20$, for which $(u_{\text{min}},u_{\text{max}})=$(0.00029, 0.051), (0.00019, 0.033), (0.0011, 0.053), (0.0023, 0.25), (0.015, 0.44), all from top to bottom. Note the typical roll structure associated to the Rayleigh-B\'enard instability, most apparent for the largest $T$.}
\label{figs75}
\end{figure}

\begin{figure}
\includegraphics[width=8.8cm,clip]{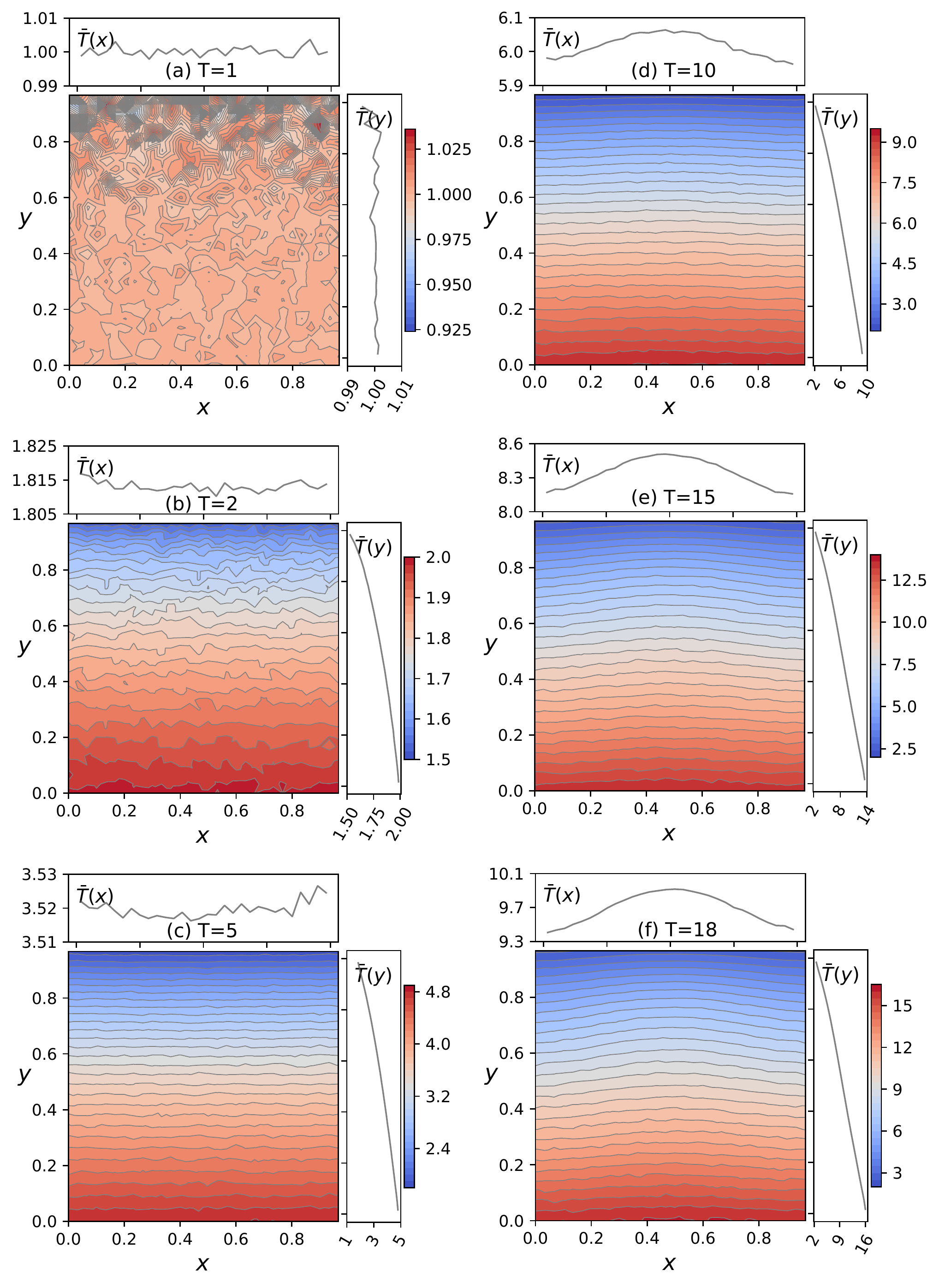}
\caption{{\bf Temperature field.} Color maps for the average temperature field $\la T(\vec{r})\ra$ measured for gravity $g=10$ and different values of the bottom plate temperature $T$, together with the average profiles across the vertical and horizontal directions. Note the nontrivial spatial structure of $\la T(\vec{r})\ra$ for strong enough temperature gradients, as well as the asymmetric, nonlinear shape of temperature profiles along the vertical direction. 
}
\label{figT}
\end{figure}

\begin{figure}
\includegraphics[width=9cm,clip]{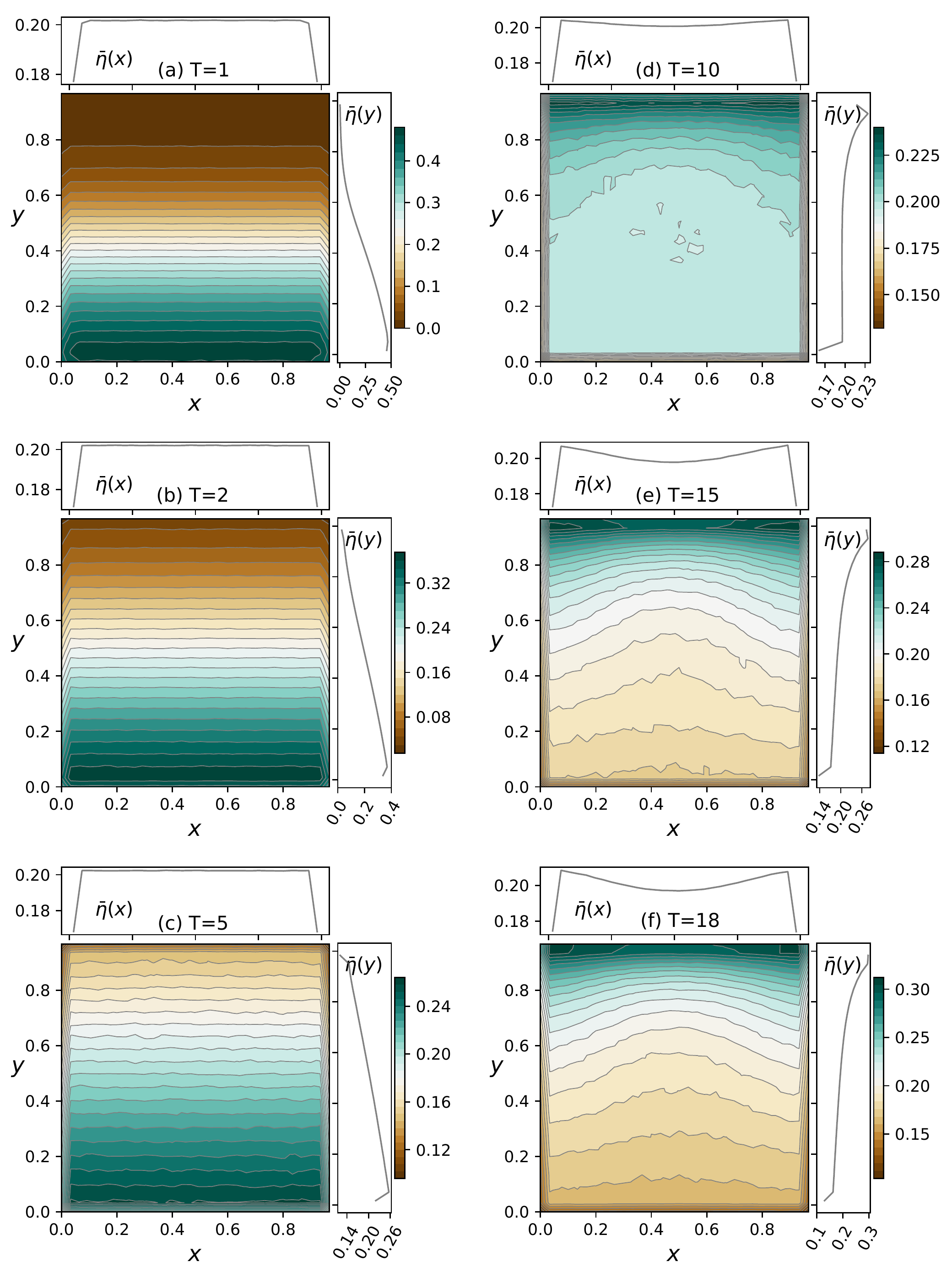}
\caption{{\bf Packing fraction field.} Color maps for the average packing fraction field $\la \eta(\vec{r})\ra$ measured for gravity $g=10$ and different values of $T$, together with the average profiles across the vertical and horizontal directions. The packing fraction field exhibits a nontrivial spatial structure, with nonlinear profiles and clear boundary effects. Note also the density gradient inversion in the vertical direction for intermediate values of $T$.
}
\label{figrho}
\end{figure}

\section{Onset of convection}

A main observable of interest in the RB instability is the average hydrodynamic velocity field $\la \vec u(\vec r)\ra$ and its emergent spatial structure. Fig.~\ref{figs75} shows $\la \vec u(\vec r)\ra$ as measured for five different bottom plate temperatures $T$ and the three values of $g>0$. First, note that the magnitude of $\la \vec u(\vec r)\ra$ is typically very small compared with the average mean particle velocity (e.g. 0.1 vs 3 for $g=10$ and $T=20$), a result of the hydrodynamic separation of scales which difficults the numerical analysis of convective structures. A clear transition from a structureless velocity field, consistent with laminar transport, to an ordered roll pattern is observed for all values of $g$ as $T$ increases. On closer inspection, however, while disorder dominates for the smallest values of $T$ considered in Fig.~\ref{figs75} $\forall g$, some incipient local order (but still roll-free) seems to emerge for intermediate values of $T$ (second row in Fig.~\ref{figs75}), though fluctuations still dominate, paving the way to fully-developed convective rolls for large enough $T$'s. Note also that convective rolls are noisier for lower values of $g$ (e.g. $g=5$ in Fig.~\ref{figs75}), in particular near the hot bottom plate. 

\begin{figure}
\includegraphics[width=8.5cm,clip]{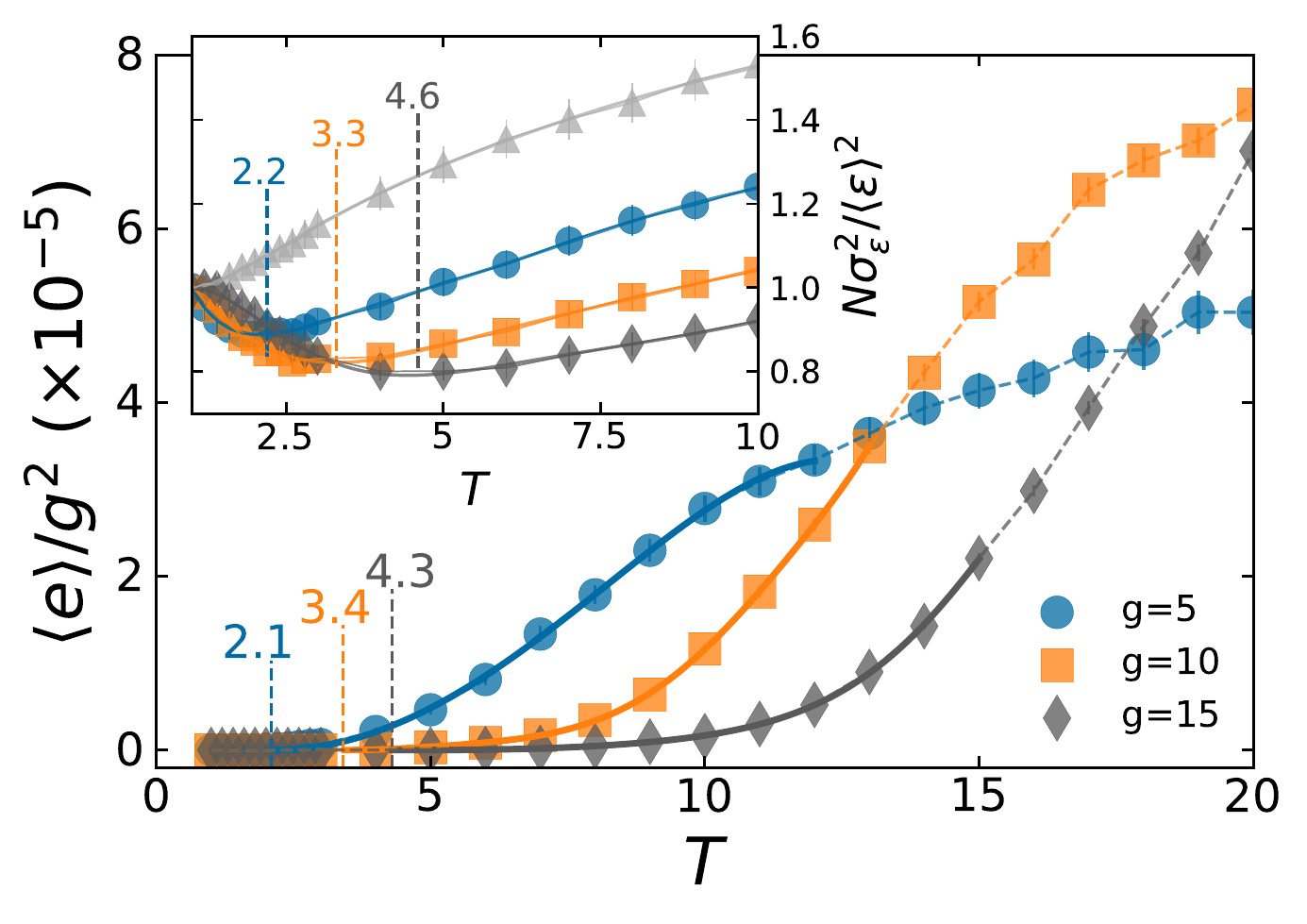}
\vspace{-0.5cm}
\caption{{\bf Order parameter.} Scaled average hydrodynamic kinetic energy, $\la e\ra/g^2$, as a function of $T$ for different values of $g$. Solid curves are polynomial fits (see text). Vertical dashed lines locate the critical temperatures obtained from these fits. Inset: Scaled relative fluctuations $N\sigma^2_\varepsilon/\la \varepsilon\ra^2$ for the kinetic energy per particle. Solid curves are fits to the data, and vertical dashed lines locate their minima. Errorbars are included in both plots.}
\label{figs73}
\end{figure}

The emergence of nontrivial structure as the bottom plate temperature increases is also apparent in other hydrodynamic fields. In particular, Fig.~\ref{figT} shows the average temperature field $\la T(\vec{r})\ra$, as well as average temperature profiles along the vertical and horizontal directions, $\bar{T}(y)$ and $\bar{T}(x)$ respectively, for a particular gravity field $g=10$ and different bottom plate temperatures, while Fig.~\ref{figrho} displays equivalent measurements for the packing fraction field $\la \eta(\vec{r})\ra$ and its vertical and horizontal profiles. Interestingly, these measured hydrodynamic fields clearly signal a strong departure from the traditional Oberbeck-Boussinesq picture for the convection instability \cite{landau13a,chandrasekhar13a,gray76a}. Indeed, temperature profiles along the vertical direction, $\bar{T}(y)$, are clearly asymmetric and nonlinear for all values of the bottom plate temperature $T$, see Fig.~\ref{figT}. Moreover, the temperature field also exhibits estructure along the $x$-direction for large enough $T$, as shown e.g. by the averaged profiles $\bar{T}(x)$. The packing fraction field also develops a nontrivial spatial structure, see Fig.~\ref{figrho}, signaling again a clear departure from the standard Oberbeck-Boussinesq theory. In particular, there is a nonlinear density gradient in the vertical direction for all values of $T$. Note also that, while high (low) packing fractions dominate near the bottom (top) plate for low $T$'s, see Figs.~\ref{figrho}.(a)-(c), this packing fraction gradient structure is reversed as $T$ grows, see Figs.~\ref{figrho}.(d)-(f), accompanied by a marked depletion effect around the central region at $x=1/2$. For completeness, Appendix \ref{appA2} contains additional data for the reduced pressure field for the same parameter points as in Figs.~\ref{figT} and \ref{figrho}.

To characterize unambiguously the emerging structure, we now define an order parameter for the RB transition. In general, we expect $\la \vec u(\vec r)\ra=0$ $\forall \vec{r}$ for non-convective states, while $\la \vec u(\vec r)\ra \neq 0$ at least in some regions once convection kicks in. This suggests using the average hydrodynamic kinetic energy $\la e\ra = \frac{\Delta^2}{2\rho}\sum_{\vec{r}} \la\eta(\vec{r})\ra \la \vec u(\vec r)\ra ^2$ as order parameter to characterize the RB instability \footnote{Previous works \cite{wesfreid78a} have used the maximum of the velocity field as order parameter. This observable is subject to stronger fluctuations, while the hydrodynamic kinetic energy has better averaging properties and allows for a more accurate characterization of the RB transition.}. Fig.~\ref{figs73} shows the measured $\la e\ra/g^2$ as a function of $T$ for all $g>0$. A first observation is that the numerical values for $\la e\ra$ are about three orders of magnitude smaller than the average kinetic energy per particle $\la \varepsilon\ra$, another fingerprint of the hydrodynamic separation of scales mentioned above. However, the large amount of data gathered allows for a significant characterization of the transition point. In particular, while $\la e\ra$ remains very close to zero for low values of $T$, there exists a nontrivial critical temperature $T_c(g)$ where $\la e\ra$ starts growing steadily, as expected for a proper order parameter. To estimate this critical temperature, we fit a piecewise-defined polynomial to the data, i.e. $\la e\ra_{\text{fit}}(T)=[T-T_c(g)]^2 \sum_{k=1}^8 a_k T^k$ for $T\ge T_c(g)$ and $\la e\ra_{\text{fit}}(T)=0$ otherwise, in the ranges $T\in[2.6,12]$, $T\in[4,13]$ and $T\in[5,16]$ for $g=5$, 10 and 15, respectively, where the fitting parameters are $T_c(g)$ and the $a_k$-coefficients, see solid curves in Fig.~\ref{figs73}. We choose a leading $[T-T_c(g)]^2$ scaling as a minimal assumption in order to have both continuity and zero first-derivative at $T=T_c(g)$, as suggested by the data. The fits are excellent in a broad region around the transition in all cases, and the critical temperatures so obtained are $T_c(g=5)=2.1$, $T_c(g=10)=3.4$ and $T_c(g=15)=4.3$. These critical temperatures, that change only slightly when varying the polynomial degree, the scaling exponent or the fitting ranges, hence signal the onset of convection for the different values of $g$ in the hard disks fluid. 

The clear-cut change of regime happening at $T_c(g)$ is also evident from the analysis of some molecular properties of the fluid. For instance, the inset to Fig.~\ref{figs73} shows the scaled relative fluctuations of the kinetic energy per particle, $N \sigma_\varepsilon^2/\la\varepsilon\ra^2$ (see Appendix \ref{appA}) as a function of $T$ $\forall g$ explored. In the absence of gravity, $g=0$, the relative kinetic energy fluctuations grow monotonically with $T$. This contrasts starkly with all $g\ne 0$ cases, for which $N \sigma_\varepsilon^2/\la\varepsilon\ra^2$ exhibits a clear minimum at a nontrivial $T$. To estimate the minima locations, we fit a generic $9$th-degree polynomial to the data  for each $g>0$ and look for the temperature where its first derivative vanishes. In this way we estimate that the minima appear at temperatures $T=2.2$, $3.3$ and $4.6$ for $g=5$, $10$ and $15$, respectively, which agree closely with the critical temperatures $T_c(g)$ derived above from hydrodynamic measurements. We hence conclude that $T_c(g)$ separates different hydrodynamic regimes in the fluid.

\begin{figure}
\includegraphics[width=8.7cm,clip]{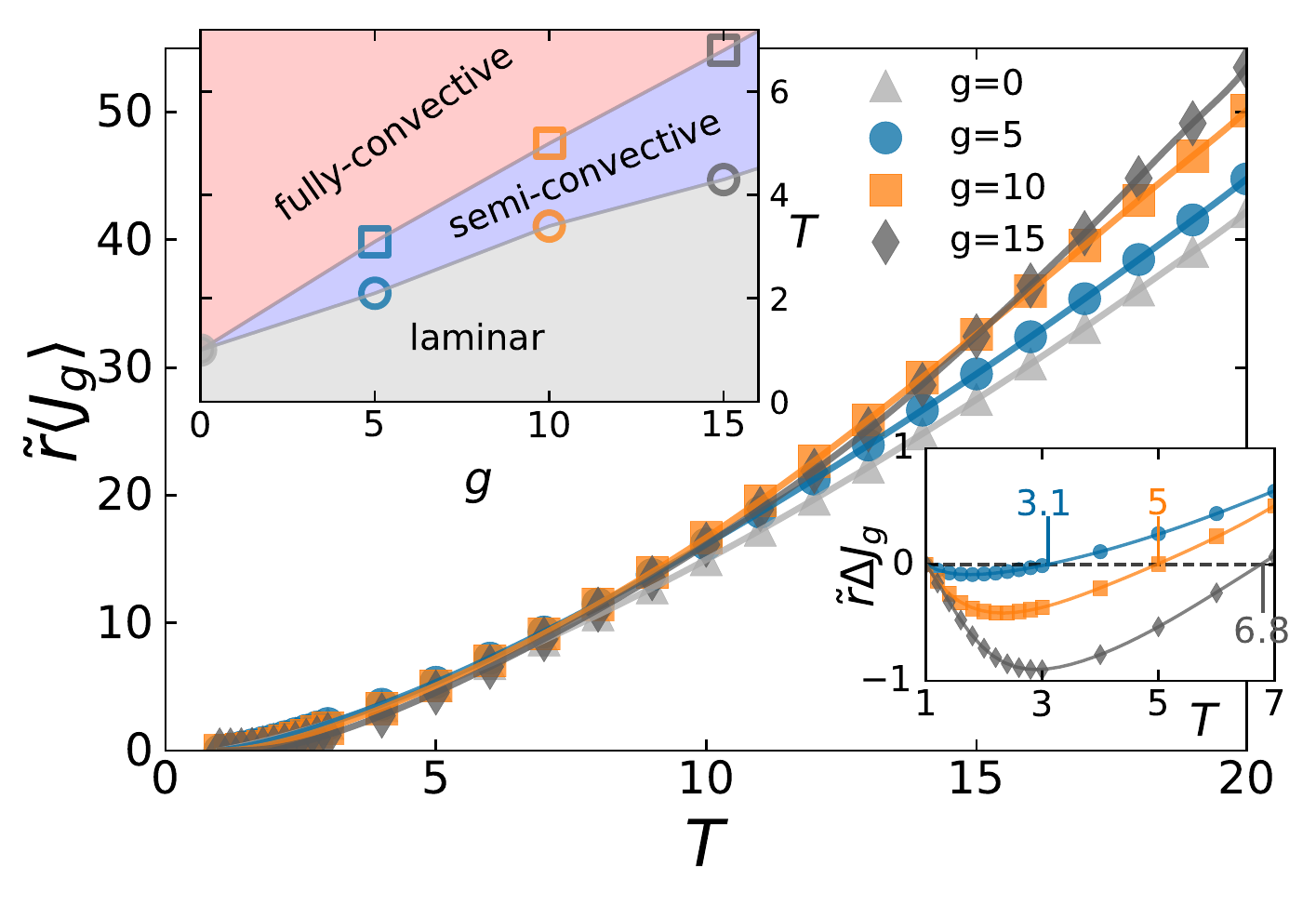}
\vspace{-0.8cm}
\caption{{\bf Convective heat transport.} Average reduced heat current $\tilde{r}\la J_g\ra$ as a function of $T$ and different $g$, with ${\tilde r}$ the disks radius. Lines are fits to the data, see text. Bottom inset: Excess current compared to the gravity-free case, $\Delta J_g = \la J_g\ra - \la J_0\ra$, as a function of $T$ and different $g\ne 0$. Vertical lines mark the temperatures $T_J(g)$ where $\Delta J_g[T_J(g)]=0$. Top inset: $T_c(g)$ ($\bigcirc$) and $T_J(g)$ ($\Box$) as a function of $g$. The different transport regimes are highlighted in color.
}
\label{figs74}
\end{figure}

\section{Convective heat transport}

Another way of studying the transition to convective flow is to characterize how the fluid's transport changes with $T$. With this aim in mind, we also measured the average heat current $\la J_g\ra$ flowing through the fluid using the energy exchange with the thermal walls during hard disks collisions (see Appendix \ref{appA}). As expected, the heat current measured at the bottom and top plates has equal magnitude and opposite sign, since no energy accumulation happens in the system. Fig.~\ref{figs74} shows $\la J_g\ra$ measured at the bottom plate as a function of $T$ for all $g$'s. In all cases $\la J_g\ra$ grows with $T$ smoothly and \emph{nonlinearly}, implying a non-constant fluid's thermal conductivity depending on the local temperature field. Moreover, $\la J_g\ra \sim T^{3/2}$ for large enough values of $T$ $\forall g$, while its initial slope for small $T$ \emph{decreases} with increasing $g$. This gives rises to a remarkable phenomenon for intermediate values of $T$ which helps us to understand the complex role of gravity in transport. In particular, all curves $\la J_g\ra$ with fixed $g>0$ go below the gravity-free curve $\la J_0\ra$ up to a nontrivial temperature $T_J(g)$ where they intersect \footnote{Note that the observation that $\la J_g\ra < \la J_0\ra$ for $1<T<T_J(g)$ implies an estimated Nusselt number $\text{Nu}\approx \la J_g\ra /\la J_0\ra<1$, see Appendix \ref{appB2}, in apparent contradiction with a bound $\text{Nu}>1$ obtained within the Oberbeck-Boussinesq hydrodynamics approximation. However, as shown in Figs.~\ref{figT} and \ref{figrho} and the associated discussion, our fluid model is clearly compressible and non-Oberbeck-Boussinesq in the parameter range studied, and hence there is no contradiction with the aforementioned bound. This is another instance of the ill-defined behavior of the dimensionless numbers of linear hydrodynamics under strong-driving conditions.}. 
This is apparent by plotting the excess current $\Delta J_g = \la J_g\ra - \la J_0\ra$, see bottom inset in Fig.~\ref{figs74}. In this way, $T_J(g)$ separates two different regimes for each $g>0$: (i) a gravity-suppressed transport regime $1<T<T_J(g)$ where gravity \emph{hinders} heat transport ($\la J_g\ra < \la J_0\ra$), and (ii) a gravity-enhanced transport regime $T>T_J(g)$ where $\la J_g\ra > \la J_0\ra$. We can measure $T_J(g)$ by fitting a curve $a_g T + b_g T^{3/2}$ to the $\la J_g\ra$ data for each $g$, with excellent results in all cases. Top inset in Fig.~\ref{figs74} shows the measured values for $T_J(g)$ ($=3.1,\, 5,\, 6.8$ for $g=5,\, 10,\,15$, respectively) as well as the critical temperatures $T_c(g)$ obtained above, and they are clearly different, $T_J(g)>T_c(g)$ $\forall g>0$. This shows that convection not always enhances heat transfer. In particular, for each $g$ there is a temperature range $T_c(g)<T<T_J(g)$ within the gravity-suppressed transport regime where convection has already kicked in (as reflected by a structured velocity field) yet it is not efficient enough to improve the transfer of energy with respect to the gravity-free ($g=0$), conductive case. We call this region $T_c(g)<T<T_J(g)$ the \emph{semi-convective regime}, see top inset in Fig.~\ref{figs74}, to distinguish it from the \emph{fully-convective regime} appearing for $T>T_J(g)$.

\begin{figure}
\includegraphics[width=8.cm,clip]{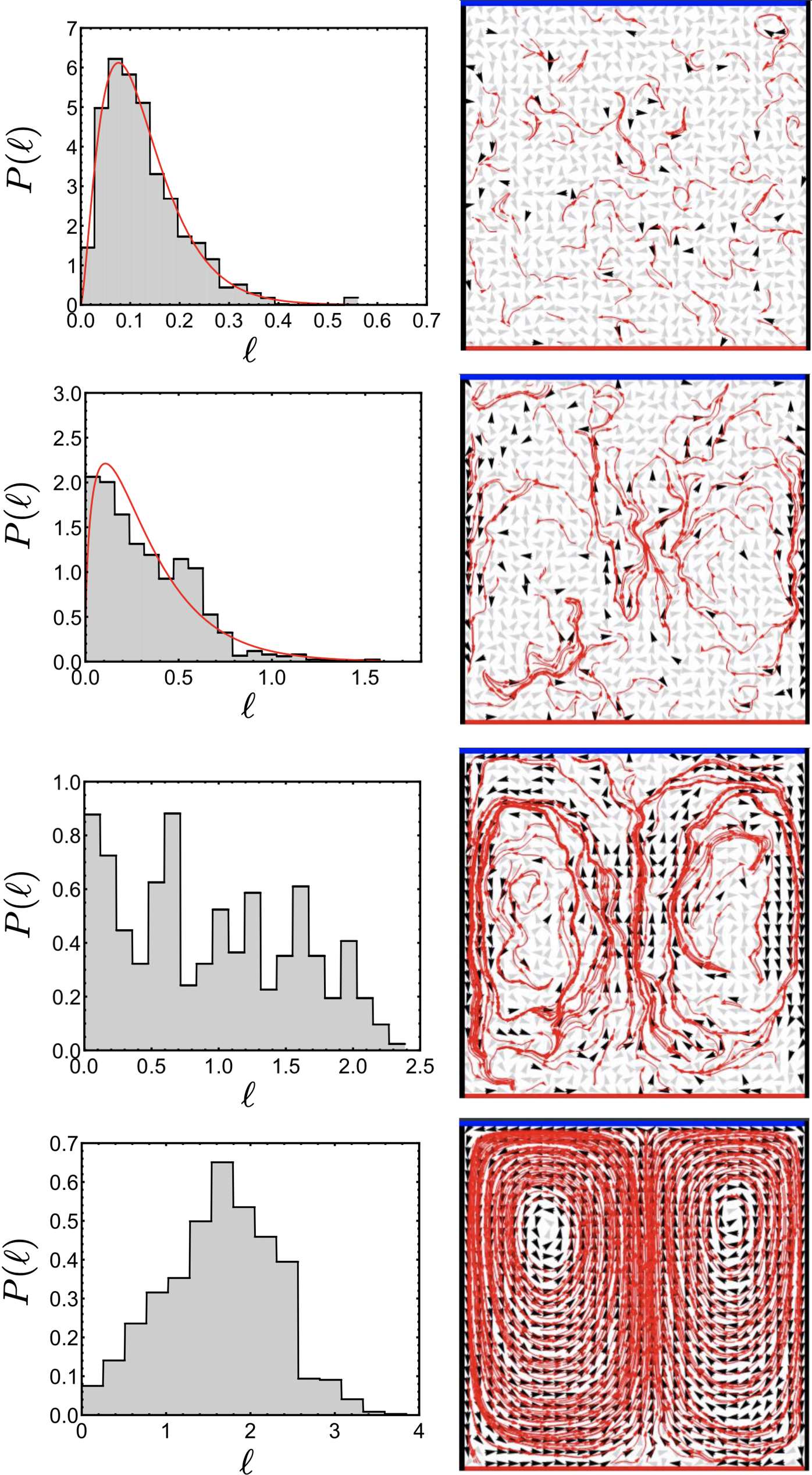}   
\caption{{\bf Active advection zones and streamlines.} Right: Average hydrodynamic velocity field $\la\vec{u}(\vec{r})\ra$ for $g=10$ and $T=1.8,\,4,\, 6, \, 20$ (from top to bottom). For $g=10$ we have $T_c=3.4$ and $T_J=5$. Active advection cells ($|\la\vec{u}(\vec{r})\ra|>\sigma[\vec{u}(\vec{r})]$) are marked with dark vectors. The red lines correspond to $10^2$ streamlines sampled from the associated $\la\vec{u}(\vec{r})\ra$. Left: Streamline length distribution in each case. Solid lines are fits to the data. 
}
\label{figs76}
\end{figure}

\section{Percolation and streamline distribution}

\begin{figure}
\includegraphics[width=8.5cm,clip]{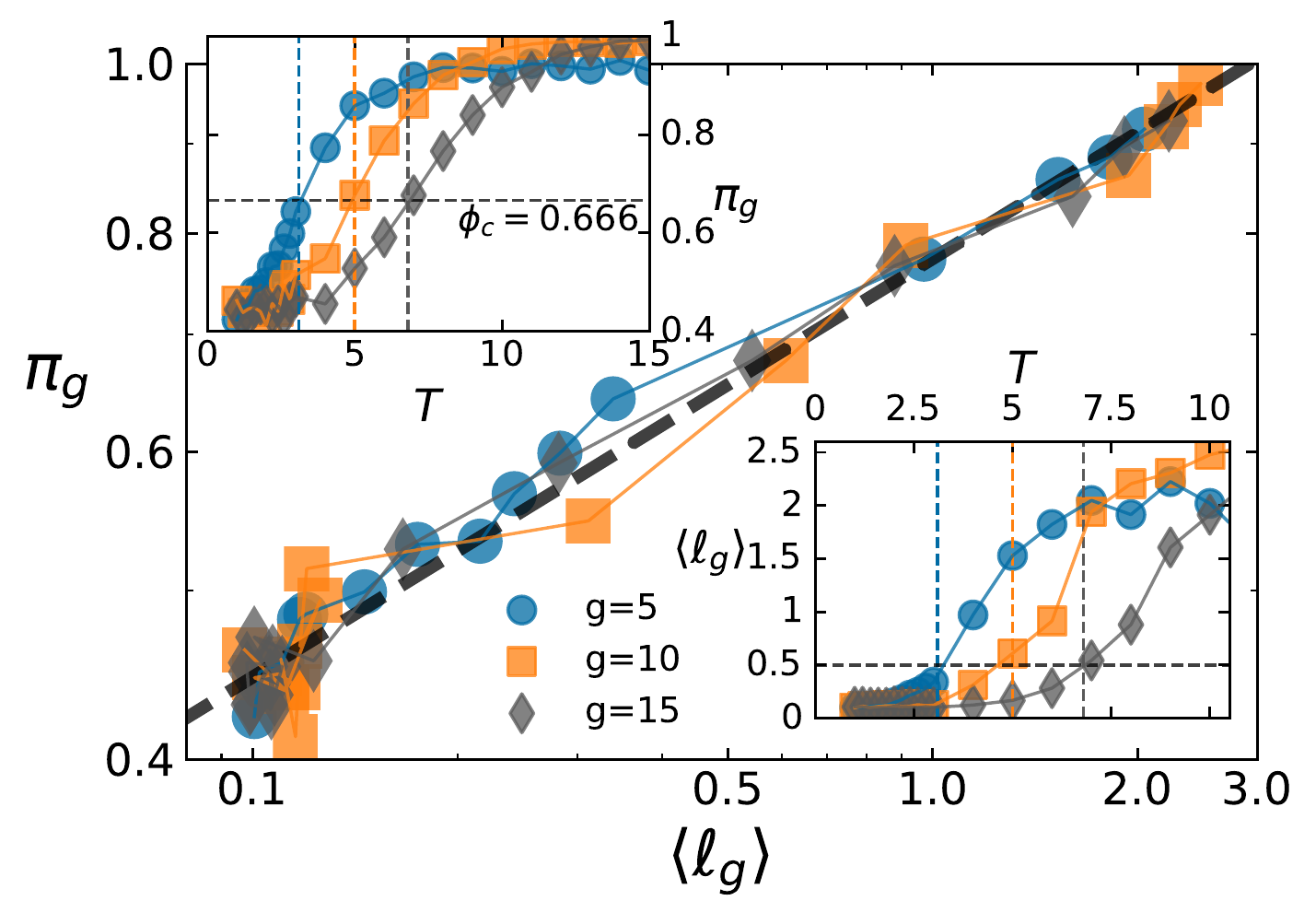}
\vspace{-0.4cm}
\caption{{\bf Percolation transition in convection.} Main: Log-log plot of the fraction $\pi_g(T)$ of active advection zones as a function of the average streamline length $\la \ell_g \ra (T)$ $\forall g>0$. Notice the striking collapse observed for all curves. The thick dashed line is a power-law $\pi_g\sim\la \ell_g\ra^{1/4}$. Top inset: $\pi_g(T)$ vs $T$ $\forall g>0$. Bottom inset: $\la \ell_g \ra$ vs $T$ $\forall g>0$. The dashed vertical lines mark $T_J(g)$ for each $g$ in both insets, while the horizontal lines signal $\la \ell_g\ra=0.5$ (bottom inset) and the critical covered area fraction $\phi_c=0.666$ (top inset).
}
\label{figs77}
\end{figure}

How can convection result in suppressed energy transport? What is the mechanism triggering efficient heat conduction? To answer these questions, we explore in more detail the structure of the average hydrodynamic velocity field $\la\vec{u}(\vec{r})\ra$ as $T$ is varied. In particular, we investigate the amount of coherent motion that a given velocity field $\la\vec{u}(\vec{r})\ra$ can sustain by measuring the number $\Pi_g(T)$ of local cells where the modulus of the local velocity vector is larger than its standard deviation, i.e. $|\la\vec{u}(\vec{r})\ra|>\sigma[\vec{u}(\vec{r})]$ (Appendix \ref{appA}), see right panels in Fig.~\ref{figs76}. This measures the number of \emph{active advection zones} in the fluid, i.e. local regions where net flow happen and hydrodynamic motion is significant against the naturally-occurring fluctuations. Top inset in Fig.~\ref{figs77} shows the fraction $\pi_g(T)=\Pi_g(T)/n_c^2$ of active advection zones, which grows monotonously with $T$ $\forall g>0$, as expected. For $T<T_J(g)$ the fraction of active advection zones is not only relatively small, but also local advection is mostly disordered, see dark arrows in the upper-right panels of Fig.~\ref{figs76}. Such disordered hydrodynamic flow is not efficient to promote heat transport against the gravitational field, and hence $\la J_g\ra < \la J_0\ra$ in this regime. However, as $T$ increases and approaches $T_J(g)$, the fraction $\pi_g(T)$ grows but so does the ratio of \emph{orientationally-aligned} active advection zones, leading to a sudden growth of the lengthscale of coherent motion, as we will see below. Most remarkably, the fraction active advection zones at the temperature $T_J(g)$ where convection transport becomes efficient takes an universal, $g$-independent value $\pi_g[T_J(g)] \approx 0.67$ $\forall g>0$, see dashed lines in the top inset of Fig.~\ref{figs77}. This value is very close to the critical covered area fraction $\phi_c\approx 0.666$ for two-dimensional continuous percolation of overlapping randomly-placed (aligned) cells \cite{mertens12a}, strongly suggesting that the transition to efficient convective heat transport is nothing but a percolation transition of active advection zones which form a spanning cluster connecting the hot and cold boundary layers, see right panels in Fig.~\ref{figs76}. Interestingly, similar percolation phenomena have been recently related with other hydrodynamic instabilities, as in e.g. Couette and pipe flows  \cite{avila11a,shi13a,lemoult16a,klotz22a}.

To quantify the range of coherent motion observed in the fluid, we now investigate the \emph{streamline statistics} for a given velocity field using integration techniques routinely used in imaging processing \cite{stalling95a,cabral93a}. In particular, given a fixed $\la\vec{u}(\vec{r})\ra$, we define a streamline starting at some initial point as the trajectory whose tangent at any point corresponds to the local velocity (see Appendix~\ref{appC}). Streamlines end whenever the underlying velocity vector field breaks the continuity of a trajectory. In this way, for a given average hydrodynamic velocity field $\la\vec{u}(\vec{r})\ra$, we generate $10^4$ streamlines with random initial points uniformly distributed in the simulation box, and compute the probability distribution $P(\ell)$ of streamline path lengths $\ell$. The right panels in Fig.~\ref{figs76} show samples of $10^2$ streamlines obtained for $g=10$ and varying $T$, and the corresponding left panels show the measured $P(\ell)$ in each case. Interestingly, $P(\ell)$ changes appreciably as we move from the non-convective to the fully-convective regime. As expected, the length distribution in the non-convective regime $T<T_c(g)$ is strongly peaked at small $\ell$ ($\sim 0.1$). In this regime $P(\ell)$ is well fitted $\forall g$ by a gamma distribution, namely $P(\ell)=A \ell^{\alpha-1}\text{e}^{-\ell/\beta}$, with $\alpha$ and $\beta$ fitting parameters. As $T$ increases beyond $T_c(g)$, $P(\ell)$ widens but maintains its gamma-shaped form, see second left panel in Fig.~\ref{figs76}. However, as we move to $T\gtrsim T_J(g)$, the length distribution broadens drastically and develops a fat tail in $\ell$, another signature of critical behavior around $T_J(g)$. This sharp change coincides with the onset of rolls in the streamline structure, see Fig.~\ref{figs76}. Streamlines capture coherent motion in the flow, and hence it is no surprise that for $T\gtrsim T_J(g)$ streamlines densely concentrate on top of the spanning clusters of aligned active advection zones, see Fig.~\ref{figs76}, associated to the roll pattern. Finally, for $T=20$ $P(\ell)$ peaks at large values of $\ell\sim 2$, exhibiting a sharp cutoff around the maximal length $\ell\sim 3$. A direct measure of the range of coherent motion in the fluid is given by the average streamline length $\la \ell_g\ra=\sum_\ell \ell \, P(\ell)$. The bottom inset in Fig.~\ref{figs77} shows the measured $\la \ell_g\ra$, which grows as a function of $T$ $\forall g>0$, and takes an universal, $g$-independent value at the onset of efficient convective transport, $\la \ell_g\ra[T_J(g)]\approx 0.5$ $\forall g>0$. This is half the system size and the lengthscale of the emerging roll structure at $T_J(g)$ connecting the bottom and top boundary layers. Indeed, plotting the fraction of active advection zones $\pi_g(T)$ against the average streamline length $\la \ell_g\ra(T)$ we find a striking collapse of data $\forall g>0$, see main panel in Fig.~\ref{figs77}. Furthermore, both observables are related via a power-law scaling $\pi_g\sim\la \ell_g\ra^{1/4}$ in a broad region around the transition temperature $T_J(g)$. Such collapse and power-law scaling are additional fingerprints of universal behavior strongly supporting the existence of an underlying percolation transition.

\section{Discussion}

In this work we have investigated at the molecular level the onset of convection in a $2d$ compressible hard disk fluid under a temperature gradient in a gravitational field, finding a surprising two-step transition scenario unobserved so far. As the bottom plate temperature increases, the fluid reaches a first critical temperature $T_c(g)$ where the hydrodynamic velocity field develops an incipient (but still roll-free) structure, and coherent local motions kick in. This first instability, apparent in the average flow field, is clearly detected both by hydrodynamic order parameters and fluid's molecular properties. The observed coherent flow is however local and disconnected, and hence unable to promote energy transfer against the gravitational field, thus resulting in inefficient heat transport. As the bottom plate temperature keeps increasing, the density and orientational order of active advection zones increase, eventually leading to a continuous percolation transition at a second critical temperature $T_J(g)>T_c(g)$ where a spanning cluster of active advection zones emerges connecting the hot and cold boundary layers and thus leading to efficient heat transport. Evidences of this percolation transition appear in the precise value of the critical covered area fraction at the transition point $\forall g>0$, but also in the streamline length distribution, which develops fat tails near $T_J(g)$, and the power-law scaling of the fraction of active advection zones in terms of the average lengthscale for coherent motion. Overall, our data strongly support the existence of two different critical temperatures $T_c(g)<T_J(g)$, with different physical origin, for the onset of convection in a two-dimensional compressible fluid. This two-step behavior seems to be linked to the compressibility of the underlying flow field, so we do not expect the two-step transition to be observable in experiments which satisfy the Oberbeck-Boussinesq approximation \cite{gray76a}.

A finite-size scaling analysis of both transitions with increasing number of particles $N$ would be of course desirable. However, going beyond $N\sim 10^3$ while reaching the massive statistics and high levels of precision needed to discriminate and fully characterize both transitions seems challenging nowadays. In any case, the bulk-boundary decoupling phenomenon reported for hard particle systems \cite{pozo15b,pozo15a,hurtado16a,hurtado20a} suggests that this two-step picture for the convection transition, that can be tested in laboratory experiments, indeed survives in the thermodynamic limit $N\to\infty$. A larger $N$ would allow us also to approach in simulations the ideal continuum limit where many particles coexist at once in a local mesoscopic cell of size $\Delta$. However, as demonstrated repeatedly in the past \cite{pozo15b,pozo15a,hurtado16a,mulero08a,szasz00a,hurtado20a}, hard particle fluids exhibit excellent self-averaging properties that, when supplemented by extensive local statistics, allow us to distill the relevant hydrodynamic fields even for moderate values of $N$, thus strongly supporting the macroscopic character of the observed behavior. Moreover, the simplicity and versatility of the two-step transition scenario here described suggests that it can also be present in more realistic models of fluid physics in three dimensions.

What happens at the hydrodynamic level in the semi-convective regime $T_c(g) < T < T_J(g)$? Our simulations support the possible existence of novel piecewise-continuous solutions to the fully-nonlinear, compressible Navier-Stokes equations, different from the well-known regular convective ones, bridging the laminar (purely conductive) fluid state for $T<T_c(g)$ with the fully-convective, rolling state for $T>T_J(g)$ once the Rayleigh instability is triggered. This possibility is worth exploring from a theoretical perspective, possibly within the fluctuating hydrodynamics framework \cite{zarate06a}, see also Refs. \cite{zaitsev71a,graham74a,swift77a}, as it challenges fundamental results in hydrodynamics, opening the door to hidden solutions and a deeper knowledge of the internal structure of Navier-Stokes equations. At the experimental level, we want to stress that small discrepancies were found in the critical Rayleigh number obtained in experiments measuring the RB transition using the velocity field as order parameter \cite{berge74a,wesfreid78a} and those obtained in experiments based on heat flux measurements (Nusselt number) \cite{behringer77a}. The chances are that such discrepancies are due to the two-step nature of the RB transition here uncovered. Moreover, experiments on fluctuations below the RB instability \cite{wu95a} have found evidences of a \emph{fluctuating flow regime} right before the instability, characterized by the appearance of fluctuating convection rolls of random orientation that resemble the coherent but disordered flow patterns that characterize the semi-convective regime here reported.

\begin{acknowledgments}
Financial support from Spanish Ministry MINECO Project No. FIS2017-84256-P and Junta de Andaluc\'{\i}a Grant No. A-FQM-175-UGR18, both supported by the European Regional Development Fund, is acknowledged. This work is also part of the Project of I+D+i Ref. PID2020-113681GB-I00, financed by MICIN/AEI/10.13039/501100011033 and FEDER \emph{A Way to Make Europe}. We are also grateful for the computational resources and assistance provided by PROTEUS, the supercomputing center of Institute Carlos I for Theoretical and Computational Physics at the University of Granada, Spain.
\end{acknowledgments}

\appendix

\section{Measurements in simulations}
\label{appA}

In this Appendix we discuss technical details on the measurement of different local observables in simulations. Similar discussion and definitions apply for global observables, so we focus only on local ones. To characterize the spatial structure of the nonequilibrium steady state, we hence measure locally different magnitudes of interest. For that, we divide the unit box ($L=1$) into $n_c\times n_c$ virtual square cells of side $\Delta=1/n_c$, and here we use $n_c=30$ to capture the fine details of the hydrodynamic fields. A given cell is characterized by a pair of integer indexes $(n_x,n_y)$, with $n_x,n_y=1,\ldots,n_c$, which correspond to a macroscopic point $\vec{r}\equiv (x,y)$ in the system, with $x,y\in[0,L=1]$, such that $x=(n_x-1/2)/n_c$ and $y=(n_y-1/2)/n_c$. In this way we use the center of each virtual cell as the macroscopic position of the local hydrodynamic fields. Hereafter we will refer indistinctly to a given cell by giving its macroscopic position $\vec{r}$ or its pair of indices $(n_x,n_y)$. Let $\bar \vec{a}(\vec r_i,\vec v_i)$ be a microscopic observable depending on a particle position $\vec r_i$ and/or velocity $\vec v_i$. This observable can be for instance the velocity of a particle $\bar \vec{a}(\vec r_i,\vec v_i)=\vec v_i$, its kinetic energy $\bar a(\vec r_i,\vec v_i)=\vec v_i^2/2$, or the local potential energy $\bar a(\vec r_i,\vec v_i)=g \, y_i$, with $\vec r_i=(x_i,y_i)$. The extensive value of this observable on cell $(n_x,n_y)$ at time $t$, or equivalently at position $\vec{r}\equiv (x,y)$, is given by
\be
A(\vec{r};t)=\sum_{j=1}^N \bar a[\vec r_j(t),\vec v_j(t)] \, \mathbb{I}_{\Omega(\vec{r})}\left[\vec r_j(t)\right] \nonumber
\label{eqapp1}
\ee
where $\Omega(\vec{r}) = \Omega(n_x,n_y)$ is the spatial domain associated to cell $(n_x,n_y)$ with center at $\vec{r}$, and $\mathbb{I}_{\Omega(\vec{r})}\left[\vec r_j(t)\right]$ is the characteristic function associated to this domain, i.e. $\mathbb{I}_{\Omega(\vec{r})}\left[\vec r_j(t)\right] = 1$ $\forall r_j(t)\in \Omega(\vec{r})$ and $\mathbb{I}_{\Omega(\vec{r})}\left[\vec r_j(t)\right] = 0$ otherwise. Note that the number of particles at cell $(n_x,n_y)$ at time $t$ is simply 
\be
N(\vec{r};t)=\sum_{j=1}^N \mathbb{I}_{\Omega(\vec{r})}\left[\vec r_j(t)\right] \, .\nonumber
\label{eqapp2}
\ee
Once the steady state has been reached in the simulation, we perform $M$ measurements of $A(\vec{r};t_k)$ at equispaced times $t_k$, $k\in[1,M]$, during the system evolution. The (extensive) average of the local observable of interest is thus trivially defined as 
\be
\la A(\vec{r})\ra = \frac{1}{M}\sum_{k=1}^M A(\vec{r};t_k) \, ,\nonumber
\label{averA}
\ee
while the (intensive) average per particle is now given by
\be
\la a(\vec{r})\ra =\frac{\la A(\vec{r})\ra}{\la N(\vec{r})\ra}  \, ,\nonumber
\label{aver}
\ee 
where we have defined the average number of particles observed in cell $\vec{r}$ as $\la N(\vec{r}) \ra= M^{-1} \sum_{k=1}^M N(\vec{r};t_k)$. Note that, when compared with the time average of $A(\vec{r};t_k)/N(\vec{r};t_k)$, the previous averaging method for intensive magnitudes exhibits a better convergence to the limiting $M\to \infty$ ensemble value $\la a\ra_\infty$, also yielding smaller fluctuations for the same $M$. 

For the error analysis, we assume that the set $\{A(\vec{r};t_k)\}_{k=1}^M$ of $M$ measurements of a given local observable is decorrelated in time so that the law of large numbers applies. In this case we expect for the average value of $A$ in the large-$M$ limit
\be
\la A(\vec{r})\ra \simeq \la A(\vec{r})\ra_\infty + \sigma[A(\vec{r})] ~\xi\nonumber
\label{averA2}
\ee
where $\xi$ is a Gaussian random variable with zero mean and unit variance, and
\be
\sigma[A(\vec{r})]  = \frac{1}{M}\sqrt{\sum_{k=1}^M\left[A(\vec{r};t_k)-\la A(\vec{r})\ra \right]^2} \nonumber
\label{sigmaA}
\ee
In this work our estimate for the asymptotic ensemble average will be $\la A(\vec{r})\ra_\infty=\la A(\vec{r})\ra \pm 3\sigma[A(\vec{r})]$, so $99.7\%$ of the data are within the errorbars shown in the analysis of the main text. Errors for intensive observables and derived magnitudes follow from the standard quadratic propagation of errors. Note also that the same discussion and definitions apply for global observables.

In this paper we are interested in a number of local hydrodynamic observables, as e.g. the average hydrodynamic velocity field $\la \vec u(\vec r)\ra$ or the packing fraction field $\la \eta(\vec r)\ra$. Following our previous discussion, $\la \vec u(\vec r)\ra$ is defined as
\be
\la \vec u(\vec r)\ra = \frac{1}{\la N(\vec r)\ra}\frac{1}{M}\sum_{k=1}^M \sum_{j=1}^N \vec v_j(t_k) \mathbb{I}_{\Omega(\vec{r})}[\vec{r}_j(t_k)] \, , \nonumber
\label{velo}
\ee
while the packing fraction field is
\be
\la \eta(\vec r)\ra = \frac{\la N(\vec{r})\ra \pi \tilde{r}^2}{\Delta^2}\nonumber
\label{pack}
\ee
where we recall that $\tilde{r}$ is the disk radius and $\Delta=1/n_c$ is the linear size of each local square cell. We will be also interested in the total hydrodynamic kinetic energy
\be
\la e\ra = \frac{\Delta^2}{2 \rho} \sum_{\vec{r}} \la \eta(\vec{r})\ra \la \vec{u}(\vec{r})\ra^2\nonumber
\label{etot}
\ee
In addition, we measure some average global molecular properties, as e.g. the average kinetic energy per particle
\be
\la \varepsilon\ra = \frac{1}{M} \sum_{k=1}^M \frac{1}{N} \sum_{j=1}^N \frac{1}{2} \vec{v}_j(t_k) \, ,\nonumber
\label{aveepsilon}
\ee
as well as its variance 
\be
\sigma^2(\varepsilon) = \la \varepsilon^2\ra - \la \varepsilon\ra^2 = \frac{1}{M} \sum_{k=1}^M \left(\frac{1}{N} \sum_{j=1}^N \frac{1}{2} \vec{v}_j(t_k)\right)^2 - \la \varepsilon\ra^2\nonumber
\ee
Finally, the energy current $\la J_g \ra$ traversing the fluid in the stationary state for gravity $g$ also plays an important role in this problem, and can be measured in each thermal wall as the accumulated sum of kinetic energy variation when disks collide with the bath wall divided by the total measurement time and the boundary length. In particular, if the steady-state simulation lasts for a total time $\tau$, and letting 
$\{i_n,t_n\}_\tau$ be the set of all particles colliding with the bottom thermal wall during this time interval $\tau$, such that particle $i_n$ collides at time $t_n$, we define the average heat current at the bottom plate for a given $g$ as
\be
\la J_g \ra = \frac{1}{2\tau L} \sum_{i_n\in\{i_n\}_\tau} [v_{i_n,y}'(t_n)^2 - v_{i_n,y}(t_n)^2] \, ,\nonumber
\ee
where $v_{i_n,y}'(t_n)$ [resp. $v_{i_n,y}(t_n)$] is the $y$-component of the velocity of disk $i_n$ right after (resp. right before) the collision with the bottom plate happening at time $t_n$.

\begin{figure}
\includegraphics[width=8.5cm,clip]{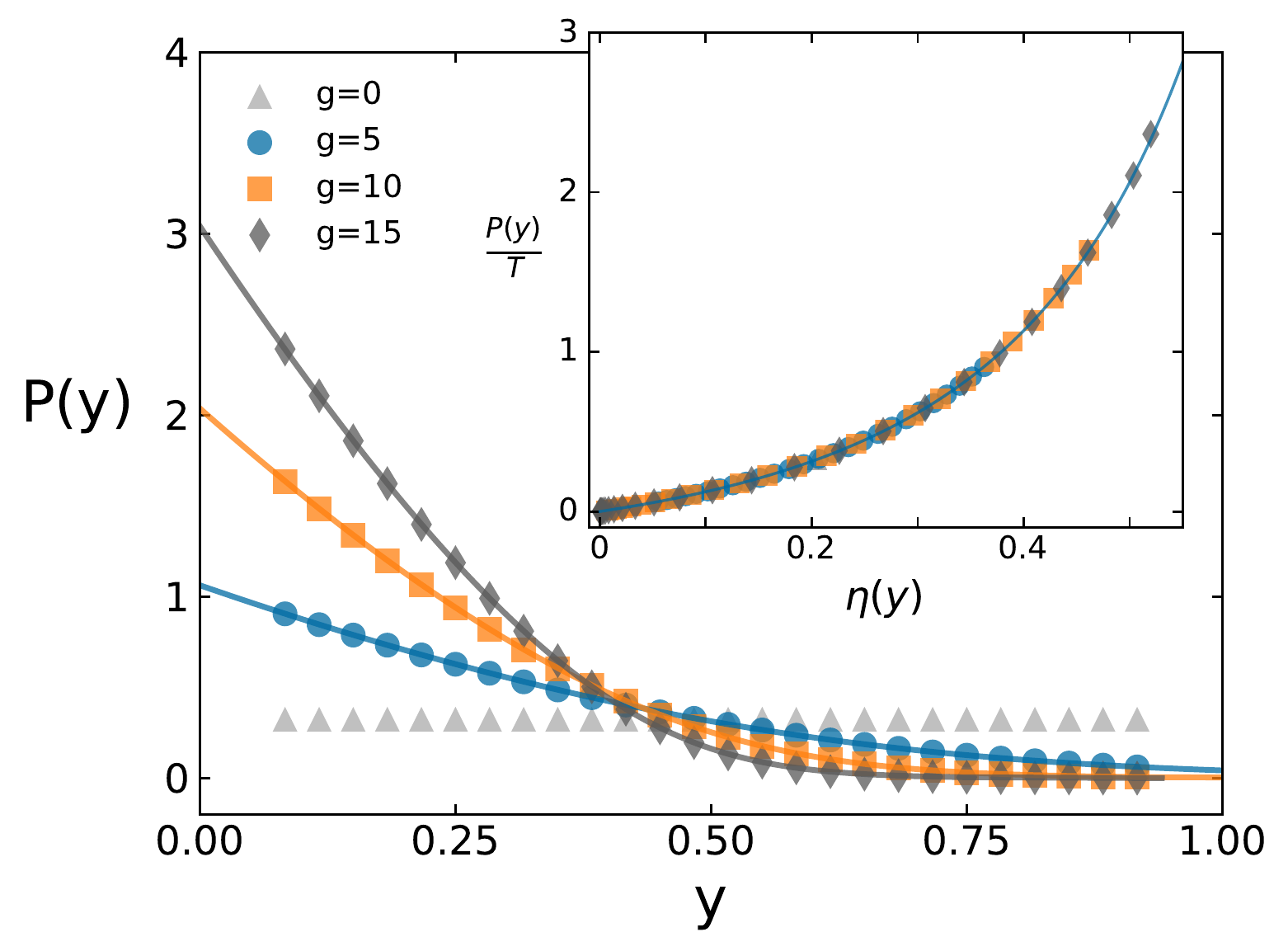}
\includegraphics[width=8.5cm,clip]{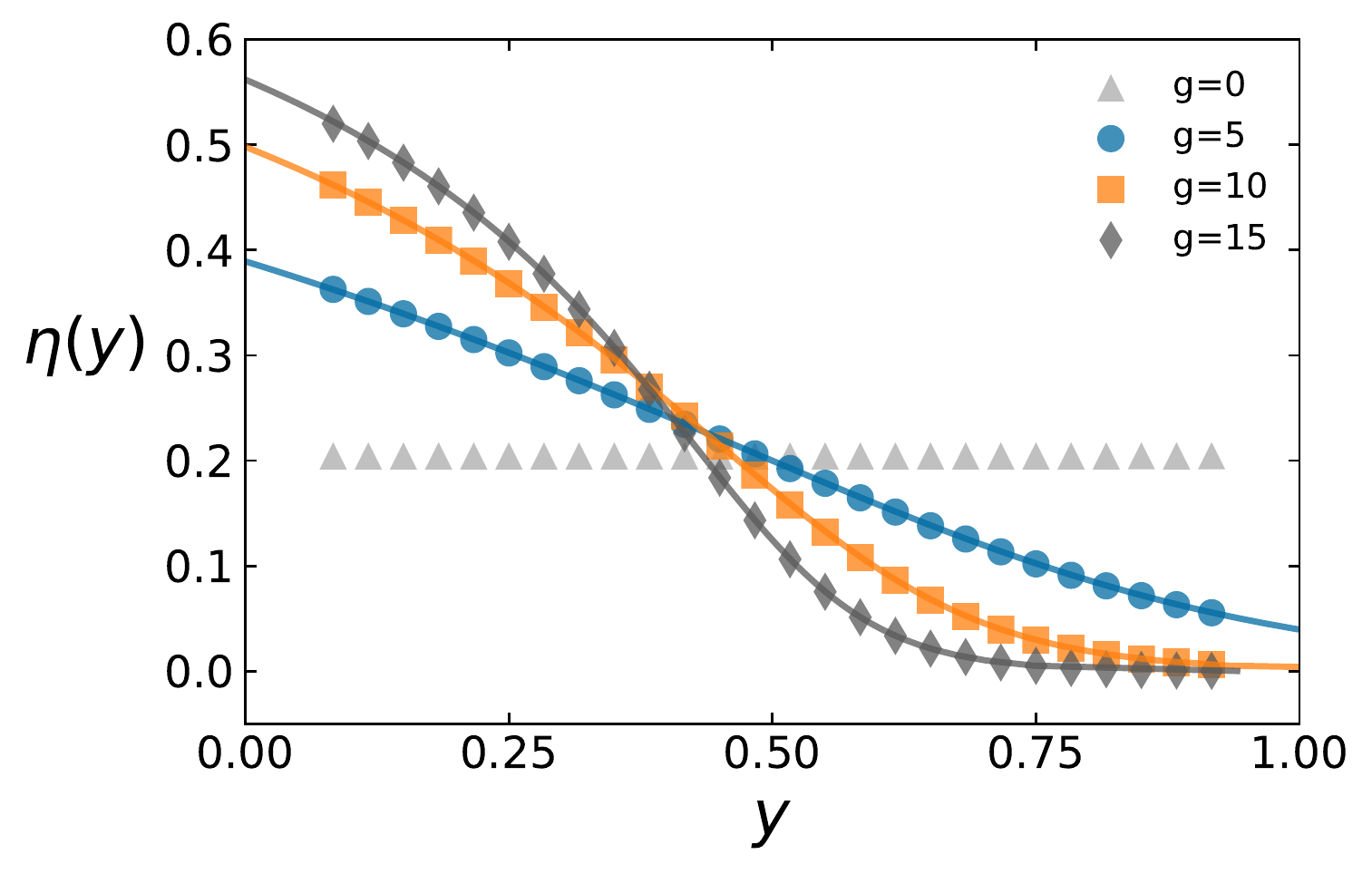}
\caption{Top: Average reduced pressure profiles measured along the vertical direction for a bottom plate temperature $T=1$ (equal to the top plate temperature) and different values of gravity $g$. Full lines correspond to the macroscopic prediction in each case. Inset: Macroscopic local equilibrium and equation of state (EoS) for hard disks. The measured local reduced pressure over the local temperature, $P(y)/T(y)$, versus the local packing fraction $\eta(y)$ for different values of $g$. The line represents the Henderson macroscopic EoS approximation \cite{henderson77a}. Bottom: Average packing fraction profiles along the vertical direction for different values of $g$, together with the macroscopic prediction derived from the hydrostatic formula and the Henderson's EoS. 
}
\label{figEq}
\end{figure}

\section{Additional data}
\label{appA2}

In this Appendix we present additional measurements which complement the results described in the main text. 

\begin{figure}
\includegraphics[width=9cm,clip]{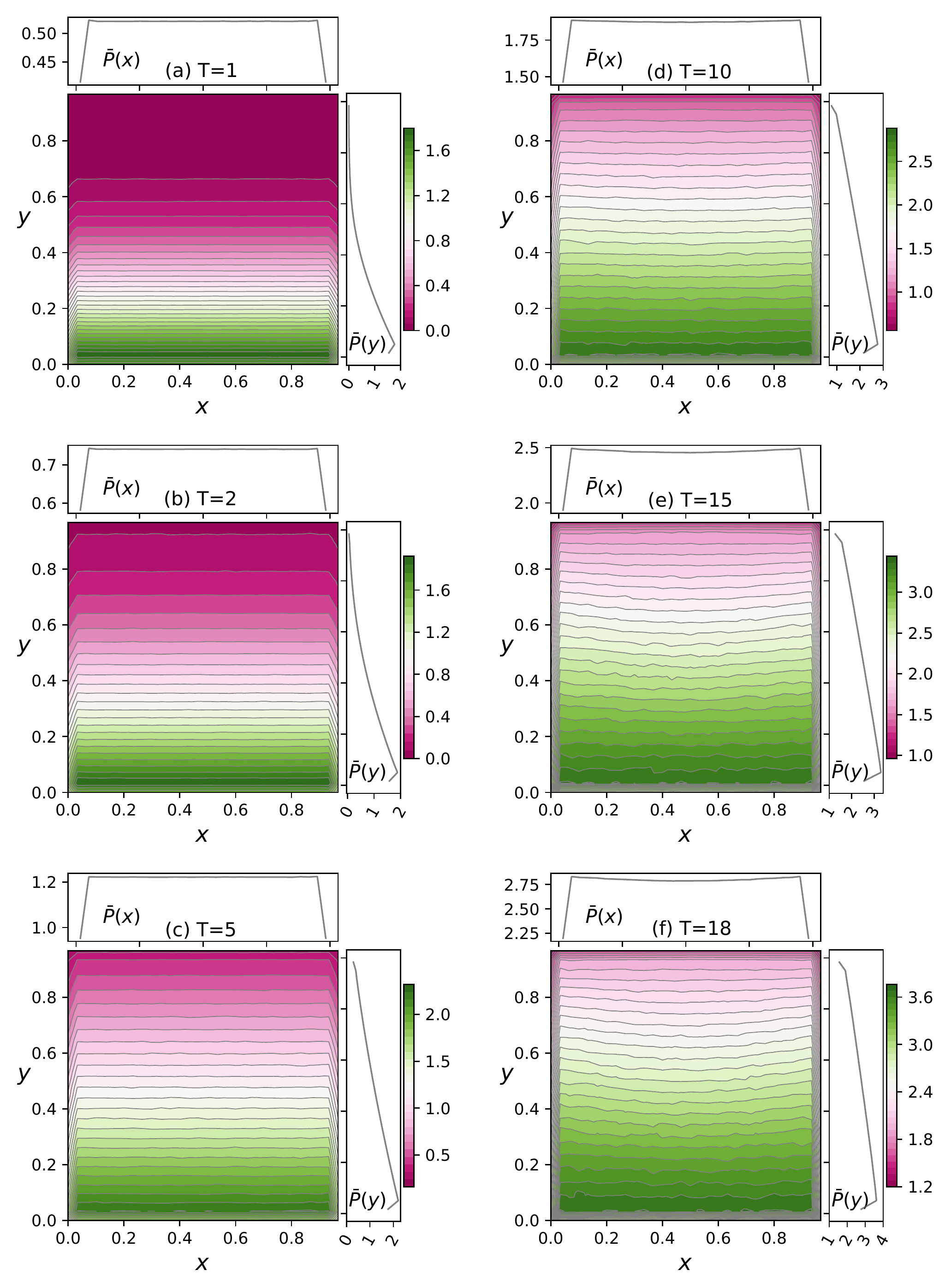}
\caption{Color maps for the reduced pressure field measured for gravity $g=10$ and different values of the bottom plate temperature $T$, together with the average profiles across the vertical and horizontal directions. 
}
\label{figP}
\end{figure}

First, to check the absence of noticeable finite-size effects in our simulations, we have measured under equilibrium conditions (equal bottom and top plate temperatures, $T=1=T_0$) both hydrostatic density and pressure profiles, as well as the local equation of state (EoS) for different gravity values and a global packing fraction $\rho=0.2$, comparing them with macroscopic formulae. In particular, the inset in the top panel of Fig. \ref{figEq} shows the measured local reduced pressure over the local temperature, $P(y)/T(y)$, versus the local average density $\rho(y)$ for different values of gravity $g$. This is nothing but a measurement of the fluid's equation of state. All data collapse onto a well-defined curve, which is captured to a high degree of accuracy by the Henderson EoS \cite{henderson77a}, demonstrating the validity of Macroscopic Local Equilibrium for the hard disks system. Note that the local average reduced pressure is defined as $P(y)=\pi\tilde{r}^2 Q(y)$, with $Q(y)$ the local average pressure at height $y$ and $\tilde{r}$ the disk radius. The main panels in Fig. \ref{figEq} show in turn the measured profiles along the vertical direction for the reduced pressure (top) and the packing fraction (bottom), together with the macroscopic predictions based on the hydrostatic formula and the Henderson EoS. In all cases the agreement between measured values and macroscopic predictions based on the continuum hypothesis are excellent.

When driven by a temperature difference in the presence of gravity, the hard disk fluid develops a non-trivial structure in its hydrodynamic fields which changes across the convection instability. In the main text we have shown how this structure evolves for the temperature and the packing fraction fields as the bottom plate temperature $T$ varies, see Figs.~\ref{figT} and \ref{figrho}. We have done this for a particular gravity field $g=10$ and different temperatures across the two-step transition. For completeness, Fig.~\ref{figP} shows the average reduced pressure field for the same parameter points as in Figs.~\ref{figT} and \ref{figrho}, as well as the associated reduced pressure profiles along the vertical and horizontal directions. The pressure field clearly exhibits a non-trivial spatial structure, mainly along the vertical direction, as well as notorious boundary effects near the confining walls.

\section{Parameter space exploration}
\label{appB}

In this Appendix we explore the parameter space for the two-dimensional hard disk fluid in a gravitational field under a vertical temperature gradient, to determine the range for $(\rho,g,T_0,T)$ where the phenomenology of interest may emerge. Here $\rho$ is the global packing fraction, $g$ is the magnitude of the gravitational field, and $T_0$ and $T$ are the temperatures of the top and bottom plates, respectively. 

First, note that the purely kinetic structure of the disks' microscopic dynamics (i.e. they do not interact except for collisions) makes it possible to fix one of the system external parameters $(T,T_0,g)$ by just applying a time rescaling without affecting the system dynamics. In other words, if we rescale time, $t=\gamma t'$, disks velocities are rescaled in turn by $v=v'/\gamma$ so, reparametrizing the temperature of the top and bottom thermal plates as $T'_{0}=\gamma^2 T_{0}$ and $T'=\gamma^2 T$, respectively, and the gravity field as $g'=\gamma^2 g$, one can prove for a fixed $\rho$ that the dynamical evolution of a system of disks with parameters $(T,T_0,g)$ is indistinguishable from that of a system with parameters $(T',T'_0,g')$. In this way one can arbitrarily choose $\gamma=1/\sqrt T_0$ in order to fix to 1 the temperature of the cold, top plate. This trick reduces the number of external control parameter to just $(\rho, g, T)$. To obtain the behavior of any observable for arbitrary values of $T_0$ one just should apply the inverse time rescaling to the dynamical variables defining the observable of interest. 

A key observable to study the Rayleigh-B{\'e}nard instability is the dimensionless Rayleigh number ($\text{Ra}$), a magnitude whose value controls the transition between the conductive and convective flow states within the Oberbeck-Boussinesq hydrodynamics approximation. The Rayleigh number is defined as the ratio of the typical timescales for diffusive and convective thermal transport, and can be written as 
\begin{equation}
\text{Ra}=\frac{\alpha g\delta T L^3}{\tilde\nu\tilde\kappa} \, ,\nonumber
\label{Ra}
\end{equation}
where $\delta T=T-T_0>0$ measures the external temperature gradient (recall we choose $T_0=1$ hereafter), $\alpha$ is the thermal expansion coefficient , $\tilde\nu=\nu/\tilde\rho$ is the kinematic viscosity with $\tilde\rho=m\rho/\pi \tilde{r}^2$ the mass density ($\tilde{r}$ is the disk radius), $\tilde\kappa=\kappa/\tilde\rho \,C_P$, and $C_P=c_p/m$ is the specific heat capacity per unit mass. For hard disks, linearizing the Navier-Stokes equations under the Oberbeck-Boussinesq approximation \cite{chandrasekhar13a}, one arrives at a critical Rayleigh number $\text{Ra}_c=27\pi^4/4\simeq 657.51$ above which convection kicks in for the stress free boundary condition case. 

\begin{figure}
\includegraphics[width=7.5cm,clip]{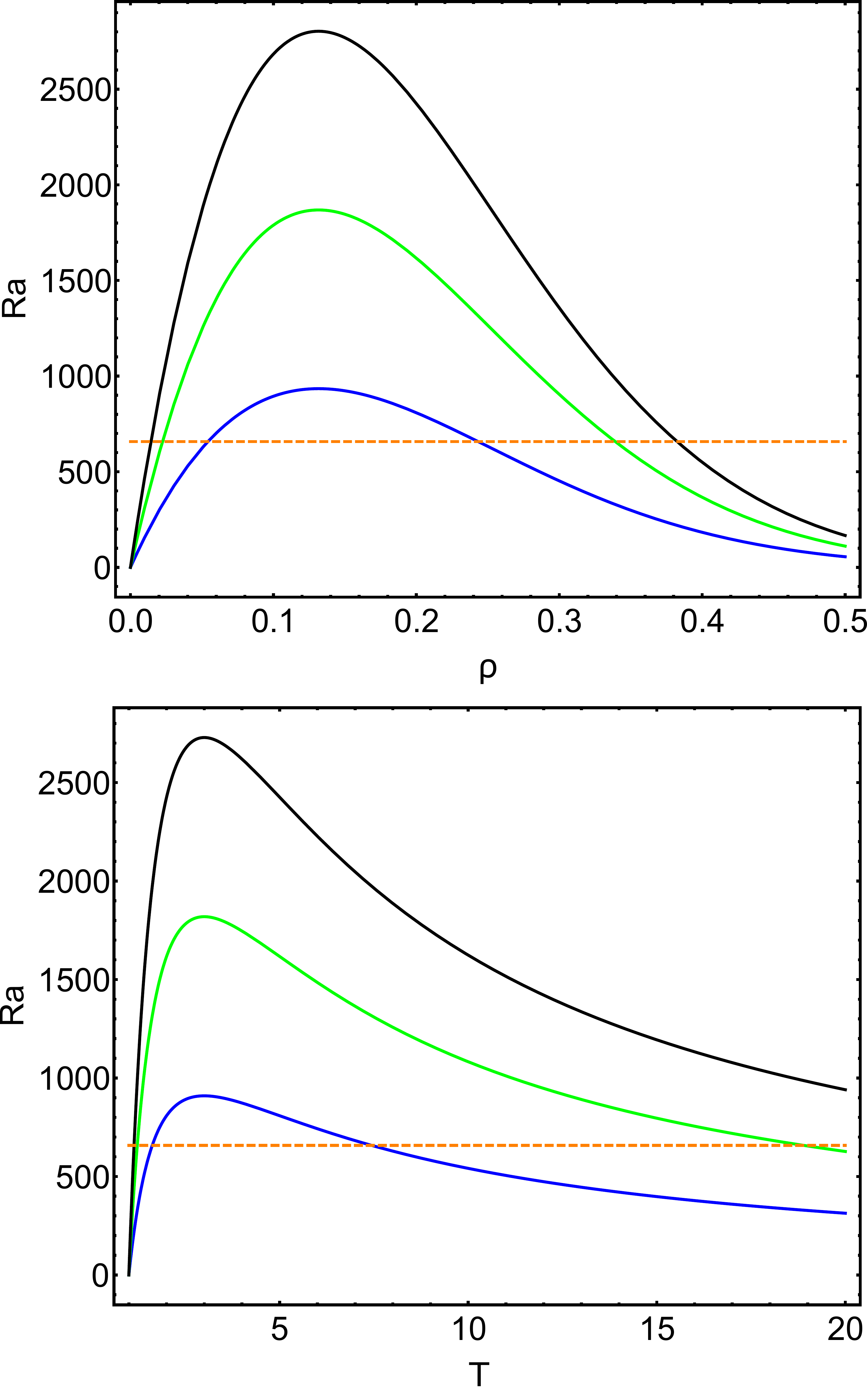}
\caption{Top: Rayleigh number $\text{Ra}$ in the Enskog approximation as a function of the global packing fraction $\rho$ for $T=5$, $T_0=1$ and $g=5$ (bottom blue curve), $g=10$ (middle green curve) and  $g=15$ (top black curve). Dashed orange line signals the critical Rayleigh number $\text{Ra}_c=27\pi^4/4\simeq 657.51$ obtained using Oberbeck-Boussinesq approximation on the Navier-Stokes equations \cite{chandrasekhar13a}. Bottom: Rayleigh number as a function of the bottom plate temperature $T$ for fixed value of $T_0=1$ and $\rho=0.2$, and for the same values of $g$ on the top panel.}
\label{figs71}
\end{figure}

To have some intuition on the range of parameters of interest, we computed $\text{Ra}$ using the Henderson equation of state approximation \cite{henderson77a} and the Enskog transport coefficients for hard disks \cite{mulero08a,hurtado20a} which work pretty well for low and moderate densities, see Fig.~\ref{figs71}. The top panel in this figure shows the behavior of $\text{Ra}$ as a function of the global packing fraction $\rho$ for $T=5$ and different values of the gravity field, namely $g=5$, $10$ and $15$. In all cases, the maximum of $\text{Ra}(\eta)$ appears for low packing fractions and it is above the critical Rayleigh number $\text{Ra}_c$. It therefore seems convenient to fix the global packing fraction to $\rho=0.2$ in our simulations, see main text. 

To gain some intuition on the value of the critical temperature $T$ for the bottom, hot plate separating non-convective and convective regimes, we plot in the bottom panel of Fig.~\ref{figs71} $\text{Ra}$ as a function of $T$ for a fixed packing fraction $\rho=0.2$. From these plots we obtain the following theoretical estimations for the critical temperatures as a function of $g$, namely $T_c(g=5)=1.6205$,  $T_c(g=10)=1.2233$ and  $T_c(g=15)=1.1376$. In this way, to cover the full range of temperatures where the interesting phenomenology emerges $\forall g>0$,
our simulations will focus on the following bottom wall temperatures across the Rayleigh-B{\'e}nard instability, namely $T=1,1.2,1.4,1.6,1.8, 2.0, 2.2, 2.4, 2.6, 2.8, 3, 4, \ldots, 19, 20$, with $g=0$, $5$, $10$ and $15$, and $\rho=0.2$, as mentioned above. 

It should be noted here that the critical temperatures $T_c(g)$ measured in simulations do not agree with the Rayleigh critical temperatures derived from hydrodynamic stability theory in this Appendix. Moreover, the theoretical analysis predicts that convection is expected to appear for lower temperatures as we increase $g$, in stark contrast with measurements (see main text), while the non-monotonic behavior of $\text{Ra}$ as a function of $T$ clearly indicates that this is not the most suitable parameter in describing the transition in the present case. The reason for these (otherwise expected) discrepancies is that the underlying theoretical approximation is linear and assumes the fluid to be incompressible, while our numerical analysis shows instead that the hard disk fluid in this strongly-driven parameter regime behaves as a fully-nonlinear and compressible (non-Oberbeck-Boussinesq) fluid \cite{bormann01a}. In any case, the analysis in this Appendix provides a first approximation to detect the range of parameters where the phenomenology of interest may emerge, and is validated \emph{a fortiori}  in our simulations.

\section{Adimensional numbers}
\label{appB2}

To facilitate the connection between our simulations and the classical hydrodynamic description of the convection instability, based on the Oberbeck-Boussinesq approximation, we now  \emph{locally} define \cite{pandey21a} some of the adimensional numbers that characterize the instability, as e.g. the Nusselt or Knudsen numbers. Note that under the strong driving conditions of these simulations the hard disk system behaves as a fully compressible fluid, with highly nonlinear hydrodynamic fields displaying a nontrivial, asymmetric structure in the gradient direction (see Figs.~\ref{figs75}-\ref{figrho} in the main text). Moreover the nonlinear dependence of the different transport coefficients on the local hydrodynamic fields becomes apparent and essential to understand the emergent macroscopic behavior. These effects signal a clear departure from the standard Oberbeck-Boussinesq approximation for the convection instability, and imply that the associated adimensional numbers (Knudsen, Nusselt, etc.) all will exhibit depth variations. Still, it seems meaningful to provide their local value at some representative state points in the laminar, semi-convective, and fully convective regimes, so as to facilitate formal comparisons with other simulations or experiments.

\begin{table}
\begin{tabular}{|c|c|c|c|}
\hline
 & $T=1.8$ & $T=4$ & $T=20$ \\ \hline \hline
\textrm{Knudsen$(y)$}   & 0.09  & 0.1   & 0.13   \\ \hline
\textrm{Nusselt$(y)$} & 1.07 & 1.07 & 1.23   \\ \hline  
\end{tabular}
\caption{Local Knudsen and Nusselt numbers, as measured near the center of the simulation box ($y=0.42$), for gravity $g=10$ and different bottom plate temperatures $T=1.8, 4$ and $20$. Note that, in all cases, the Knudsen number is around 0.1, meaning that the flow field is that of a continuum fluid. On the other hand, Nusselt numbers at the center of the simulation box are all above 1.}
\label{tab1}
\end{table}

We hence define the local Nusselt number at height $y$ as $\textrm{Nu}(y)=\la J_g\ra/ J_c(y)$, where $\la J_g\ra$ is the total energy current measured across the vertical walls of the simulation box capturing convective heat transfer, and $J_c(y)$ is the local conduction current at depth $y$, defined as $J_c(y) = |\kappa[\eta(y),T(y)] T'(y)|$, where $T'(y)$ is the measured local temperature gradient along the vertical direction and $\kappa[\eta(y),T(y)]$ is the Enskog kinetic theory expression for the thermal conductivity of hard disks for local packing fraction $\eta(y)$ and temperature $T(y)$ \cite{mulero08a,hurtado20a,cordero97a}. The local packing fraction and temperature correspond to those measured in simulations in each case. On the other hand, the local Knudsen number is defined as $\textrm{Knu}(y) = \lambda(y)/L$, where $L$ is the system linear size ($L=1$ in our units), and $\lambda(y)$ is the molecular mean free path as obtained from Enskog kinetic theory for values of the local packing fraction $\eta(y)$ and temperature $T(y)$ measured in simulations \cite{cordero97a}. Table~\ref{tab1} reports Nusselt and Knudsen numbers measured locally near the center of the simulation box ($y=0.42$) for gravity $g=10$ and different bottom plate temperatures $T=1.8$ (laminar flow), $T=4$ (semi-convective flow) and $T=20$ (fully-convective flow), see also top inset to Fig.~\ref{figs74} in the main text.

In all cases, the locally-measured Knudsen number is around 0.1, meaning that the associated flow is similar to that of a continuum fluid phase, and hence we can safely say that our simulations are close to the continuum limit and that we are indeed observing a continuum phenomenon. On the other hand, the local Nusselt number near the center is slightly above 1 in all cases, see Table~\ref{tab1}. This local number hence seems compatible with the bound $\textrm{Nu}>1$  on Nusselt number for incompresible fluids \cite{landau13a,behringer77a}, though our model fluid is compressible in this regime. Note however that both the local Knudsen and Nusselt numbers are expected to exhibit depth variations as a reflection of the nonlinear, asymmetric character of temperature and packing fraction profiles along the $y$-direction, see Figs.~\ref{figT}-\ref{figrho} in the main text.

\begin{table}
\begin{tabular}{|c|c|c|c|c|c|c|}
\hline
\multicolumn{1}{|c|}{} & \multicolumn{2}{|c|}{$T=1.8$} & \multicolumn{2}{|c|}{$T=4$} & \multicolumn{2}{|c|}{$T=20$} \\ \hline \hline
 & \makecell{\textrm{top}} & \makecell{\textrm{center}}  & \makecell{\textrm{top}} & \makecell{\textrm{center}} & \makecell{\textrm{top}}  & \makecell{\textrm{center}} \\ \hline
\textrm{Nusselt$(y)$} & \makecell{\textrm{0.88}} & \makecell{\textrm{1.07}} & \makecell{\textrm{0.98}} & \makecell{\textrm{1.07}} & \makecell{\textrm{1.09}} & \makecell{\textrm{1.23}}   \\ \hline  
\end{tabular}
\caption{Local Nusselt number, as measured near the top plate ($y=0.88$) and near the center of the simulation box ($y=0.42$), for gravity $g=10$ and different bottom plate temperatures $T=1.8, 4$ and $20$. Note the values of Nusselt number smaller than 1 near the top plate for $T=1.8$ and $4$.}
\label{tab2}
\end{table}

To confirm these depth variations, we also measured the local Nusselt number near the top plate ($y=0.88$) for the three representative cases mentioned above, see Table~\ref{tab2}. Interestingly, we observe local Nusselt numbers below 1 for $T=1.8$ and 4 near the top plate, in a spatial region where packing fraction is low for these values of $T$, see Fig.~\ref{figrho} in the main text. This result is fully consistent with the observation of a gravity-suppressed transport regime in both the laminar and semi-convective regimes, see Fig.~\ref{figs74} in the main text. Moreover, this local measurements confirm that the Nusselt number is not bounded to be larger than 1 in our compressible, non-Oberbeck-Boussinesq fluid.

\section{Streamline computation}
\label{appC}

In this Appendix we describe the procedure to generate streamlines from the measured flow field using integration techniques routinely used in imaging processing \cite{stalling95a,cabral93a}. For a given velocity field $\vec{u}(\vec{r})$, we define a streamline starting at some initial point as the trajectory whose tangent at any point corresponds to the given fixed vector field. That is, let $\vec u(x,y)=(u_1(x,y),u_2(x,y))$ be a fixed vector field  at a point $\vec{r}=(x,y)$, and $\vec{r}_0=(x_0,y_0)$ an arbitrary initial point. Then, the streamline associated to this point is the solution of the following differential equation
\begin{equation}
\frac{dy}{dx}=\frac{u_2(x,y)}{u_1(x,y)} \, ,\nonumber
\label{streamline1}
\end{equation}
or in parametric form
\begin{equation}
\frac{dx}{ds}=u_1(x(s),y(s))\, ,\qquad \frac{dy}{ds}=u_2(x(s),y(s)) \, ,\nonumber
\label{streamline2}
\end{equation}
with initial condition at $(x_0,y_0)$. Numerical solutions to this problem can be simply found using e.g. a standard Runge-Kutta integrator \cite{press96a}. Note that in our case the underlying velocity vector field is defined over a discrete $n_c\times n_c$ grid, with $n_c=30$, so in order to reconstruct the vector field at arbitrary points in the plane as needed by the streamline numerical integrator we perform linear interpolations from neighboring grid sites. Streamlines end whenever the underlying velocity vector field breaks the continuity of a trajectory. This can be detected by a simple stop condition in the above integration scheme: a streamline ends whenever the condition $\vec u(n+1)\cdot\vec u(n)<0$ is satisfied, where $\vec u(n)=[u_1(x(s_n),y(s_n)),u_2(x(s_n),y(s_n))]$ is the velocity vector in the $n$-th step of the integration scheme. Note that similar integration techniques to obtain streamlines from discrete vector fields are routinely used in imaging processing, see for instance Refs. \cite{stalling95a,cabral93a} for some other technicalities.


\begin{thebibliography}{74}%
\makeatletter
\providecommand \@ifxundefined [1]{%
 \@ifx{#1\undefined}
}%
\providecommand \@ifnum [1]{%
 \ifnum #1\expandafter \@firstoftwo
 \else \expandafter \@secondoftwo
 \fi
}%
\providecommand \@ifx [1]{%
 \ifx #1\expandafter \@firstoftwo
 \else \expandafter \@secondoftwo
 \fi
}%
\providecommand \natexlab [1]{#1}%
\providecommand \enquote  [1]{``#1''}%
\providecommand \bibnamefont  [1]{#1}%
\providecommand \bibfnamefont [1]{#1}%
\providecommand \citenamefont [1]{#1}%
\providecommand \href@noop [0]{\@secondoftwo}%
\providecommand \href [0]{\begingroup \@sanitize@url \@href}%
\providecommand \@href[1]{\@@startlink{#1}\@@href}%
\providecommand \@@href[1]{\endgroup#1\@@endlink}%
\providecommand \@sanitize@url [0]{\catcode `\\12\catcode `\$12\catcode
  `\&12\catcode `\#12\catcode `\^12\catcode `\_12\catcode `\%12\relax}%
\providecommand \@@startlink[1]{}%
\providecommand \@@endlink[0]{}%
\providecommand \url  [0]{\begingroup\@sanitize@url \@url }%
\providecommand \@url [1]{\endgroup\@href {#1}{\urlprefix }}%
\providecommand \urlprefix  [0]{URL }%
\providecommand \Eprint [0]{\href }%
\providecommand \doibase [0]{http://dx.doi.org/}%
\providecommand \selectlanguage [0]{\@gobble}%
\providecommand \bibinfo  [0]{\@secondoftwo}%
\providecommand \bibfield  [0]{\@secondoftwo}%
\providecommand \translation [1]{[#1]}%
\providecommand \BibitemOpen [0]{}%
\providecommand \bibitemStop [0]{}%
\providecommand \bibitemNoStop [0]{.\EOS\space}%
\providecommand \EOS [0]{\spacefactor3000\relax}%
\providecommand \BibitemShut  [1]{\csname bibitem#1\endcsname}%
\let\auto@bib@innerbib\@empty
\bibitem [{\citenamefont {Bodenschatz}\ \emph {et~al.}(2000)\citenamefont
  {Bodenschatz}, \citenamefont {Pesch},\ and\ \citenamefont
  {Ahlers}}]{bodenschatz00a}%
  \BibitemOpen
  \bibfield  {author} {\bibinfo {author} {\bibfnamefont {E.}~\bibnamefont
  {Bodenschatz}}, \bibinfo {author} {\bibfnamefont {W.}~\bibnamefont {Pesch}},
  \ and\ \bibinfo {author} {\bibfnamefont {G.}~\bibnamefont {Ahlers}},\
  }\bibfield  {title} {\enquote {\bibinfo {title} {Recent developments in
  {Rayleigh-B{\'e}nard} convection},}\ }\href {\doibase
  10.1146/annurev.fluid.32.1.709} {\bibfield  {journal} {\bibinfo  {journal}
  {Annu. Rev. Fluid Mech.}\ }\textbf {\bibinfo {volume} {32}},\ \bibinfo
  {pages} {709} (\bibinfo {year} {2000})}\BibitemShut {NoStop}%
\bibitem [{\citenamefont {Ahlers}\ \emph {et~al.}(2009)\citenamefont {Ahlers},
  \citenamefont {Grossmann},\ and\ \citenamefont {Lohse}}]{ahlers09a}%
  \BibitemOpen
  \bibfield  {author} {\bibinfo {author} {\bibfnamefont {G.}~\bibnamefont
  {Ahlers}}, \bibinfo {author} {\bibfnamefont {S.}~\bibnamefont {Grossmann}}, \
  and\ \bibinfo {author} {\bibfnamefont {D.}~\bibnamefont {Lohse}},\ }\bibfield
   {title} {\enquote {\bibinfo {title} {Heat transfer and large scale dynamics
  in turbulent {Rayleigh-B{\'e}nard} convection},}\ }\href {\doibase
  10.1103/RevModPhys.81.503} {\bibfield  {journal} {\bibinfo  {journal} {Rev.
  Mod. Phys.}\ }\textbf {\bibinfo {volume} {81}},\ \bibinfo {pages} {503}
  (\bibinfo {year} {2009})}\BibitemShut {NoStop}%
\bibitem [{\citenamefont {Mutabazi}\ \emph {et~al.}(2010)\citenamefont
  {Mutabazi}, \citenamefont {Wesfreid},\ and\ \citenamefont
  {Guyon}}]{mutabazi10a}%
  \BibitemOpen
  \bibfield  {author} {\bibinfo {author} {\bibfnamefont {I.}~\bibnamefont
  {Mutabazi}}, \bibinfo {author} {\bibfnamefont {J.~E.}\ \bibnamefont
  {Wesfreid}}, \ and\ \bibinfo {author} {\bibfnamefont {E.}~\bibnamefont
  {Guyon}},\ }\href {https://link.springer.com/book/10.1007%2Fb106790} {\emph
  {\bibinfo {title} {Dynamics of spatio-temporal cellular structures: {Henri
  B{\'e}nard} centenary review}}},\ Vol.\ \bibinfo {volume} {207}\ (\bibinfo
  {publisher} {Springer, New York},\ \bibinfo {year} {2010})\BibitemShut {NoStop}%
\bibitem [{\citenamefont {Ogura}\ and\ \citenamefont
  {Phillips}(1962)}]{ogura62a}%
  \BibitemOpen
  \bibfield  {author} {\bibinfo {author} {\bibfnamefont {Y.}~\bibnamefont
  {Ogura}}\ and\ \bibinfo {author} {\bibfnamefont {N.~A.}\ \bibnamefont
  {Phillips}},\ }\bibfield  {title} {\enquote {\bibinfo {title} {Scale analysis
  of deep and shallow convection in the atmosphere},}\ }\href {\doibase
  10.1175/1520-0469(1962)019<0173:SAODAS>2.0.CO;2} {\bibfield  {journal}
  {\bibinfo  {journal} {J. Atmos. Sci.}\ }\textbf {\bibinfo {volume} {19}},\
  \bibinfo {pages} {173} (\bibinfo {year} {1962})}\BibitemShut {NoStop}%
\bibitem [{\citenamefont {Gierasch}\ \emph {et~al.}(2000)\citenamefont
  {Gierasch}, \citenamefont {Ingersoll}, \citenamefont {Banfield},
  \citenamefont {Ewald}, \citenamefont {Helfenstein}, \citenamefont
  {Simon-Miller}, \citenamefont {Vasavada}, \citenamefont {Breneman},
  \citenamefont {Senske},\ and\ \citenamefont {Team}}]{gierasch00a}%
  \BibitemOpen
  \bibfield  {author} {\bibinfo {author} {\bibfnamefont {P.~J.}\ \bibnamefont
  {Gierasch}}, \bibinfo {author} {\bibfnamefont {A.~P.}\ \bibnamefont
  {Ingersoll}}, \bibinfo {author} {\bibfnamefont {D.}~\bibnamefont {Banfield}},
  \bibinfo {author} {\bibfnamefont {S.~P.}\ \bibnamefont {Ewald}}, \bibinfo
  {author} {\bibfnamefont {P.}~\bibnamefont {Helfenstein}}, \bibinfo {author}
  {\bibfnamefont {A.}~\bibnamefont {Simon-Miller}}, \bibinfo {author}
  {\bibfnamefont {A.}~\bibnamefont {Vasavada}}, \bibinfo {author}
  {\bibfnamefont {H.~H.}\ \bibnamefont {Breneman}}, \bibinfo {author}
  {\bibfnamefont {D.~A.}\ \bibnamefont {Senske}}, \ and\ \bibinfo {author}
  {\bibfnamefont {Galileo~Imaging}\ \bibnamefont {Team}},\ }\bibfield  {title}
  {\enquote {\bibinfo {title} {Observation of moist convection in {Jupiter's}
  atmosphere},}\ }\href {\doibase 10.1038/35001017} {\bibfield  {journal}
  {\bibinfo  {journal} {Nature}\ }\textbf {\bibinfo {volume} {403}},\ \bibinfo
  {pages} {628} (\bibinfo {year} {2000})}\BibitemShut {NoStop}%
\bibitem [{\citenamefont {Marshall}\ and\ \citenamefont
  {Schott}(1999)}]{marshall99a}%
  \BibitemOpen
  \bibfield  {author} {\bibinfo {author} {\bibfnamefont {J.}~\bibnamefont
  {Marshall}}\ and\ \bibinfo {author} {\bibfnamefont {F.}~\bibnamefont
  {Schott}},\ }\bibfield  {title} {\enquote {\bibinfo {title} {Open-ocean
  convection: Observations, theory, and models},}\ }\href {\doibase
  10.1029/98RG02739} {\bibfield  {journal} {\bibinfo  {journal} {Rev.
  Geophys.}\ }\textbf {\bibinfo {volume} {37}},\ \bibinfo {pages} {1} (\bibinfo
  {year} {1999})}\BibitemShut {NoStop}%
\bibitem [{\citenamefont {Morgan}(1971)}]{morgan71a}%
  \BibitemOpen
  \bibfield  {author} {\bibinfo {author} {\bibfnamefont {W.~J.}\ \bibnamefont
  {Morgan}},\ }\bibfield  {title} {\enquote {\bibinfo {title} {Convection
  plumes in the lower mantle},}\ }\href {\doibase 10.1038/230042a0} {\bibfield
  {journal} {\bibinfo  {journal} {Nature}\ }\textbf {\bibinfo {volume} {230}},\
  \bibinfo {pages} {42} (\bibinfo {year} {1971})}\BibitemShut {NoStop}%
\bibitem [{\citenamefont {Davies}\ and\ \citenamefont
  {Richards}(1992)}]{davies92a}%
  \BibitemOpen
  \bibfield  {author} {\bibinfo {author} {\bibfnamefont {G.~F.}\ \bibnamefont
  {Davies}}\ and\ \bibinfo {author} {\bibfnamefont {M.~A.}\ \bibnamefont
  {Richards}},\ }\bibfield  {title} {\enquote {\bibinfo {title} {Mantle
  convection},}\ }\href {\doibase 10.1086/629582} {\bibfield  {journal}
  {\bibinfo  {journal} {J. Geol.}\ }\textbf {\bibinfo {volume} {100}},\
  \bibinfo {pages} {151} (\bibinfo {year} {1992})}\BibitemShut {NoStop}%
\bibitem [{\citenamefont {Roxburgh}(1978)}]{roxburgh78a}%
  \BibitemOpen
  \bibfield  {author} {\bibinfo {author} {\bibfnamefont {I.W.}\ \bibnamefont
  {Roxburgh}},\ }\bibfield  {title} {\enquote {\bibinfo {title} {Convection and
  stellar structure},}\ }\href
  {http://adsabs.harvard.edu/full/1978A%26A....65..281R} {\bibfield  {journal}
  {\bibinfo  {journal} {Astron. Astrophys.}\ }\textbf {\bibinfo {volume}
  {65}},\ \bibinfo {pages} {281} (\bibinfo {year} {1978})}\BibitemShut
  {NoStop}%
\bibitem [{\citenamefont {Meakin}\ and\ \citenamefont
  {Arnett}(2007)}]{meakin07a}%
  \BibitemOpen
  \bibfield  {author} {\bibinfo {author} {\bibfnamefont {C.~A.}\ \bibnamefont
  {Meakin}}\ and\ \bibinfo {author} {\bibfnamefont {D.}~\bibnamefont
  {Arnett}},\ }\bibfield  {title} {\enquote {\bibinfo {title} {Turbulent
  convection in stellar interiors. i. hydrodynamic simulation},}\ }\href
  {https://iopscience.iop.org/article/10.1086/520318/meta} {\bibfield
  {journal} {\bibinfo  {journal} {Astrophys. J.}\ }\textbf {\bibinfo {volume}
  {667}},\ \bibinfo {pages} {448} (\bibinfo {year} {2007})}\BibitemShut
  {NoStop}%
\bibitem [{\citenamefont {Schumacher}\ and\ \citenamefont
  {Sreenivasan}(2020)}]{schumacher20a}%
  \BibitemOpen
  \bibfield  {author} {\bibinfo {author} {\bibfnamefont {J.}~\bibnamefont
  {Schumacher}}\ and\ \bibinfo {author} {\bibfnamefont {K.~R.}\ \bibnamefont
  {Sreenivasan}},\ }\bibfield  {title} {\enquote {\bibinfo {title} {Unusual
  dynamics of convection in the {Sun}},}\ }\href {\doibase
  10.1103/RevModPhys.92.041001} {\bibfield  {journal} {\bibinfo  {journal}
  {Rev. Mod. Phys.}\ }\textbf {\bibinfo {volume} {92}},\ \bibinfo {pages}
  {041001} (\bibinfo {year} {2020})}\BibitemShut {NoStop}%
\bibitem [{\citenamefont {Murdoch}\ \emph {et~al.}(2013)\citenamefont
  {Murdoch}, \citenamefont {Rozitis}, \citenamefont {Nordstrom}, \citenamefont
  {Green}, \citenamefont {Michel}, \citenamefont {de~Lophem},\ and\
  \citenamefont {Losert}}]{murdoch13a}%
  \BibitemOpen
  \bibfield  {author} {\bibinfo {author} {\bibfnamefont {N.}~\bibnamefont
  {Murdoch}}, \bibinfo {author} {\bibfnamefont {B.}~\bibnamefont {Rozitis}},
  \bibinfo {author} {\bibfnamefont {K.}~\bibnamefont {Nordstrom}}, \bibinfo
  {author} {\bibfnamefont {S.~F.}\ \bibnamefont {Green}}, \bibinfo {author}
  {\bibfnamefont {P.}~\bibnamefont {Michel}}, \bibinfo {author} {\bibfnamefont
  {T.-L.}\ \bibnamefont {de~Lophem}}, \ and\ \bibinfo {author} {\bibfnamefont
  {W.}~\bibnamefont {Losert}},\ }\bibfield  {title} {\enquote {\bibinfo {title}
  {Granular convection in microgravity},}\ }\href {\doibase
  10.1103/PhysRevLett.110.018307} {\bibfield  {journal} {\bibinfo  {journal}
  {Phys. Rev. Lett.}\ }\textbf {\bibinfo {volume} {110}},\ \bibinfo {pages}
  {018307} (\bibinfo {year} {2013})}\BibitemShut {NoStop}%
\bibitem [{\citenamefont {Pontuale}\ \emph {et~al.}(2016)\citenamefont
  {Pontuale}, \citenamefont {Gnoli}, \citenamefont {Vega~Reyes},\ and\
  \citenamefont {Puglisi}}]{pontuale16a}%
  \BibitemOpen
  \bibfield  {author} {\bibinfo {author} {\bibfnamefont {G.}~\bibnamefont
  {Pontuale}}, \bibinfo {author} {\bibfnamefont {A.}~\bibnamefont {Gnoli}},
  \bibinfo {author} {\bibfnamefont {F.}~\bibnamefont {Vega~Reyes}}, \ and\
  \bibinfo {author} {\bibfnamefont {A.}~\bibnamefont {Puglisi}},\ }\bibfield
  {title} {\enquote {\bibinfo {title} {Thermal convection in granular gases
  with dissipative lateral walls},}\ }\href {\doibase
  10.1103/PhysRevLett.117.098006} {\bibfield  {journal} {\bibinfo  {journal}
  {Phys. Rev. Lett.}\ }\textbf {\bibinfo {volume} {117}},\ \bibinfo {pages}
  {098006} (\bibinfo {year} {2016})}\BibitemShut {NoStop}%
\bibitem [{\citenamefont {Bar-Cohen}\ \emph {et~al.}(2007)\citenamefont
  {Bar-Cohen}, \citenamefont {Wang},\ and\ \citenamefont
  {Rahim}}]{bar-cohen07a}%
  \BibitemOpen
  \bibfield  {author} {\bibinfo {author} {\bibfnamefont {A.}~\bibnamefont
  {Bar-Cohen}}, \bibinfo {author} {\bibfnamefont {P.}~\bibnamefont {Wang}}, \
  and\ \bibinfo {author} {\bibfnamefont {E.}~\bibnamefont {Rahim}},\ }\bibfield
   {title} {\enquote {\bibinfo {title} {Thermal management of high heat flux
  nanoelectronic chips},}\ }\href {\doibase 10.1007/BF02915748} {\bibfield
  {journal} {\bibinfo  {journal} {Microgravity Sci. and Tech.}\ }\textbf
  {\bibinfo {volume} {19}},\ \bibinfo {pages} {48} (\bibinfo {year}
  {2007})}\BibitemShut {NoStop}%
\bibitem [{\citenamefont {Iyengar}\ \emph {et~al.}(2014)\citenamefont
  {Iyengar}, \citenamefont {Geisler},\ and\ \citenamefont
  {Sammakia}}]{iyengar14a}%
  \BibitemOpen
  \bibfield  {author} {\bibinfo {author} {\bibfnamefont {M.}~\bibnamefont
  {Iyengar}}, \bibinfo {author} {\bibfnamefont {K.~J.~L.}\ \bibnamefont
  {Geisler}}, \ and\ \bibinfo {author} {\bibfnamefont {B.}~\bibnamefont
  {Sammakia}},\ }\href {\doibase 10.1142/9067} {\emph {\bibinfo {title}
  {Cooling of Microelectronic and Nanoelectronic Equipment}}}\ (\bibinfo
  {publisher} {World Scientific, Singapore},\ \bibinfo {year} {2014})\BibitemShut
  {NoStop}%
\bibitem [{\citenamefont {Yu}\ \emph {et~al.}(2010)\citenamefont {Yu},
  \citenamefont {Lee},\ and\ \citenamefont {Yook}}]{yu10a}%
  \BibitemOpen
  \bibfield  {author} {\bibinfo {author} {\bibfnamefont {S.-H.}\ \bibnamefont
  {Yu}}, \bibinfo {author} {\bibfnamefont {K.-S.}\ \bibnamefont {Lee}}, \ and\
  \bibinfo {author} {\bibfnamefont {S.-J.}\ \bibnamefont {Yook}},\ }\bibfield
  {title} {\enquote {\bibinfo {title} {Natural convection around a radial heat
  sink},}\ }\href {\doibase 10.1016/j.ijheatmasstransfer.2010.02.032}
  {\bibfield  {journal} {\bibinfo  {journal} {Int. J. Heat Mass Transf.}\
  }\textbf {\bibinfo {volume} {53}},\ \bibinfo {pages} {2935} (\bibinfo {year}
  {2010})}\BibitemShut {NoStop}%
\bibitem [{\citenamefont {Kalteh}\ \emph {et~al.}(2012)\citenamefont {Kalteh},
  \citenamefont {Abbassi}, \citenamefont {Saffar-Avval}, \citenamefont
  {Frijns}, \citenamefont {Darhuber},\ and\ \citenamefont
  {Harting}}]{kalteh12a}%
  \BibitemOpen
  \bibfield  {author} {\bibinfo {author} {\bibfnamefont {M.}~\bibnamefont
  {Kalteh}}, \bibinfo {author} {\bibfnamefont {A.}~\bibnamefont {Abbassi}},
  \bibinfo {author} {\bibfnamefont {M.}~\bibnamefont {Saffar-Avval}}, \bibinfo
  {author} {\bibfnamefont {A.}~\bibnamefont {Frijns}}, \bibinfo {author}
  {\bibfnamefont {A.}~\bibnamefont {Darhuber}}, \ and\ \bibinfo {author}
  {\bibfnamefont {J.}~\bibnamefont {Harting}},\ }\bibfield  {title} {\enquote
  {\bibinfo {title} {Experimental and numerical investigation of nanofluid
  forced convection inside a wide microchannel heat sink},}\ }\href {\doibase
  10.1016/j.applthermaleng.2011.10.023} {\bibfield  {journal} {\bibinfo
  {journal} {Appl. Therm. Eng.}\ }\textbf {\bibinfo {volume} {36}},\ \bibinfo
  {pages} {260} (\bibinfo {year} {2012})}\BibitemShut {NoStop}%
\bibitem [{\citenamefont {Teertstra}\ \emph {et~al.}(2000)\citenamefont
  {Teertstra}, \citenamefont {Yovanovich},\ and\ \citenamefont
  {Culham}}]{teertstra00a}%
  \BibitemOpen
  \bibfield  {author} {\bibinfo {author} {\bibfnamefont {P.}~\bibnamefont
  {Teertstra}}, \bibinfo {author} {\bibfnamefont {M.~M.}\ \bibnamefont
  {Yovanovich}}, \ and\ \bibinfo {author} {\bibfnamefont {J.~R.}\ \bibnamefont
  {Culham}},\ }\bibfield  {title} {\enquote {\bibinfo {title} {Analytical
  forced convection modeling of plate fin heat sinks},}\ }\href {\doibase
  10.1142/S0960313100000320} {\bibfield  {journal} {\bibinfo  {journal} {J.
  Electron. Manuf.}\ }\textbf {\bibinfo {volume} {10}},\ \bibinfo {pages} {253}
  (\bibinfo {year} {2000})}\BibitemShut {NoStop}%
\bibitem [{\citenamefont {Verboven}\ \emph {et~al.}(2000)\citenamefont
  {Verboven}, \citenamefont {Scheerlinck}, \citenamefont {De~Baerdemaeker},\
  and\ \citenamefont {Nicola\"{\i}}}]{verboven00a}%
  \BibitemOpen
  \bibfield  {author} {\bibinfo {author} {\bibfnamefont {P.}~\bibnamefont
  {Verboven}}, \bibinfo {author} {\bibfnamefont {N.}~\bibnamefont
  {Scheerlinck}}, \bibinfo {author} {\bibfnamefont {J.}~\bibnamefont
  {De~Baerdemaeker}}, \ and\ \bibinfo {author} {\bibfnamefont {B.~M.}\
  \bibnamefont {Nicola\"{\i}}},\ }\bibfield  {title} {\enquote {\bibinfo
  {title} {Computational fluid dynamics modelling and validation of the
  temperature distribution in a forced convection oven},}\ }\href {\doibase
  10.1016/S0260-8774(99)00133-8} {\bibfield  {journal} {\bibinfo  {journal} {J.
  Food Eng.}\ }\textbf {\bibinfo {volume} {43}},\ \bibinfo {pages} {61}
  (\bibinfo {year} {2000})}\BibitemShut {NoStop}%
\bibitem [{\citenamefont {Calmidi}\ and\ \citenamefont
  {Mahajan}(2000)}]{calmidi00a}%
  \BibitemOpen
  \bibfield  {author} {\bibinfo {author} {\bibfnamefont {V.~V.}\ \bibnamefont
  {Calmidi}}\ and\ \bibinfo {author} {\bibfnamefont {R.~L.}\ \bibnamefont
  {Mahajan}},\ }\bibfield  {title} {\enquote {\bibinfo {title} {Forced
  convection in high porosity metal foams},}\ }\href {\doibase
  10.1115/1.1287793} {\bibfield  {journal} {\bibinfo  {journal} {J. Heat
  Transfer}\ }\textbf {\bibinfo {volume} {122}},\ \bibinfo {pages} {557}
  (\bibinfo {year} {2000})}\BibitemShut {NoStop}%
\bibitem [{\citenamefont {Brent}\ \emph {et~al.}(1988)\citenamefont {Brent},
  \citenamefont {Voller},\ and\ \citenamefont {Reid}}]{brent88a}%
  \BibitemOpen
  \bibfield  {author} {\bibinfo {author} {\bibfnamefont {A.~D.}\ \bibnamefont
  {Brent}}, \bibinfo {author} {\bibfnamefont {V.~R.}\ \bibnamefont {Voller}}, \
  and\ \bibinfo {author} {\bibfnamefont {K.~T.~J.}\ \bibnamefont {Reid}},\
  }\bibfield  {title} {\enquote {\bibinfo {title} {Enthalpy-porosity technique
  for modeling convection-diffusion phase change: application to the melting of
  a pure metal},}\ }\href {\doibase 10.1080/10407788808913615} {\bibfield
  {journal} {\bibinfo  {journal} {Numer. Heat Transfer}\ }\textbf {\bibinfo
  {volume} {13}},\ \bibinfo {pages} {297} (\bibinfo {year} {1988})}\BibitemShut
  {NoStop}%
\bibitem [{\citenamefont {Chandrasekhar}(2013)}]{chandrasekhar13a}%
  \BibitemOpen
  \bibfield  {author} {\bibinfo {author} {\bibfnamefont {S.}~\bibnamefont
  {Chandrasekhar}},\ }\href
  {https://books.google.es/books?hl=es&lr=&id=Mg3CAgAAQBAJ&oi=fnd&pg=PP1&dq=Chandrasekhar,+S.,+Hydrodynamic+and+Hydromagnetic+stability,+Dover,+New+York+(1981)&ots=cpqdmWLi91&sig=fRtEdMK5xTlISg1RG51W0yjMcl0&redir_esc=y}
  {\emph {\bibinfo {title} {Hydrodynamic and hydromagnetic stability}}}\
  (\bibinfo  {publisher} {Dover Publications, New York},\ \bibinfo {year}
  {2013})\BibitemShut {NoStop}%
\bibitem [{\citenamefont {Getling}(1998)}]{getling98a}%
  \BibitemOpen
  \bibfield  {author} {\bibinfo {author} {\bibfnamefont {A.~V.}\ \bibnamefont
  {Getling}},\ }\href
  {https://books.google.es/books?hl=es&lr=&id=a_43hQr33HcC&oi=fnd&pg=PA1&dq=Rayleigh-Be%CC%81nard+Convection:+Structures+and+Dynamics&ots=GyQ6VICEI9&sig=01ocaoy8COdRI2FEjzpJ9NTIVqA#v=onepage&q=Rayleigh-Be%CC%81nard%20Convection%3A%20Structures%20and%20Dynamics&f=false}
  {\emph {\bibinfo {title} {{Rayleigh-B{\'e}nard} Convection: Structures and
  Dynamics}}},\ Vol.~\bibinfo {volume} {11}\ (\bibinfo  {publisher} {World
  Scientific, Singapore},\ \bibinfo {year} {1998})\BibitemShut {NoStop}%
\bibitem [{\citenamefont {Brown}\ \emph {et~al.}(2005)\citenamefont {Brown},
  \citenamefont {Nikolaenko},\ and\ \citenamefont {Ahlers}}]{brown05a}%
  \BibitemOpen
  \bibfield  {author} {\bibinfo {author} {\bibfnamefont {E.}~\bibnamefont
  {Brown}}, \bibinfo {author} {\bibfnamefont {A.}~\bibnamefont {Nikolaenko}}, \
  and\ \bibinfo {author} {\bibfnamefont {G.}~\bibnamefont {Ahlers}},\
  }\bibfield  {title} {\enquote {\bibinfo {title} {Reorientation of the
  large-scale circulation in turbulent {Rayleigh-B{\'e}nard} convection},}\
  }\href {\doibase 10.1103/PhysRevLett.95.084503} {\bibfield  {journal}
  {\bibinfo  {journal} {Phys. Rev. Lett.}\ }\textbf {\bibinfo {volume} {95}},\
  \bibinfo {pages} {084503} (\bibinfo {year} {2005})}\BibitemShut {NoStop}%
\bibitem [{\citenamefont {Brown}\ and\ \citenamefont
  {Ahlers}(2007)}]{brown07a}%
  \BibitemOpen
  \bibfield  {author} {\bibinfo {author} {\bibfnamefont {E.}~\bibnamefont
  {Brown}}\ and\ \bibinfo {author} {\bibfnamefont {G.}~\bibnamefont {Ahlers}},\
  }\bibfield  {title} {\enquote {\bibinfo {title} {Large-scale circulation
  model for turbulent {Rayleigh-B{\'e}nard} convection},}\ }\href {\doibase
  10.1103/PhysRevLett.98.134501} {\bibfield  {journal} {\bibinfo  {journal}
  {Phys. Rev. Lett.}\ }\textbf {\bibinfo {volume} {98}},\ \bibinfo {pages}
  {134501} (\bibinfo {year} {2007})}\BibitemShut {NoStop}%
\bibitem [{\citenamefont {Pandey}\ \emph {et~al.}(2018)\citenamefont {Pandey},
  \citenamefont {Scheel},\ and\ \citenamefont {Schumacher}}]{pandey18a}%
  \BibitemOpen
  \bibfield  {author} {\bibinfo {author} {\bibfnamefont {A.}~\bibnamefont
  {Pandey}}, \bibinfo {author} {\bibfnamefont {J.~D.}\ \bibnamefont {Scheel}},
  \ and\ \bibinfo {author} {\bibfnamefont {J.}~\bibnamefont {Schumacher}},\
  }\bibfield  {title} {\enquote {\bibinfo {title} {Turbulent superstructures in
  {Rayleigh-B{\'e}nard} convection},}\ }\href {\doibase
  10.1038/s41467-018-04478-0} {\bibfield  {journal} {\bibinfo  {journal}
  {Nature Comm.}\ }\textbf {\bibinfo {volume} {9}},\ \bibinfo {pages} {1}
  (\bibinfo {year} {2018})}\BibitemShut {NoStop}%
\bibitem [{\citenamefont {Hartlep}\ \emph {et~al.}(2003)\citenamefont
  {Hartlep}, \citenamefont {Tilgner},\ and\ \citenamefont
  {Busse}}]{hartlep03a}%
  \BibitemOpen
  \bibfield  {author} {\bibinfo {author} {\bibfnamefont {T.}~\bibnamefont
  {Hartlep}}, \bibinfo {author} {\bibfnamefont {A.}~\bibnamefont {Tilgner}}, \
  and\ \bibinfo {author} {\bibfnamefont {F.~H.}\ \bibnamefont {Busse}},\
  }\bibfield  {title} {\enquote {\bibinfo {title} {Large scale structures in
  {Rayleigh-B{\'e}nard} convection at high {Rayleigh} numbers},}\ }\href
  {\doibase 10.1103/PhysRevLett.91.064501} {\bibfield  {journal} {\bibinfo
  {journal} {Phys. Rev. Lett.}\ }\textbf {\bibinfo {volume} {91}},\ \bibinfo
  {pages} {064501} (\bibinfo {year} {2003})}\BibitemShut {NoStop}%
\bibitem [{\citenamefont {Wang}\ \emph {et~al.}(2020)\citenamefont {Wang},
  \citenamefont {Verzicco}, \citenamefont {Lohse},\ and\ \citenamefont
  {Shishkina}}]{wang20a}%
  \BibitemOpen
  \bibfield  {author} {\bibinfo {author} {\bibfnamefont {Q.}~\bibnamefont
  {Wang}}, \bibinfo {author} {\bibfnamefont {R.}~\bibnamefont {Verzicco}},
  \bibinfo {author} {\bibfnamefont {D.}~\bibnamefont {Lohse}}, \ and\ \bibinfo
  {author} {\bibfnamefont {O.}~\bibnamefont {Shishkina}},\ }\bibfield  {title}
  {\enquote {\bibinfo {title} {Multiple states in turbulent large-aspect ratio
  thermal convection: What determines the number of convection rolls?}}\ }\href
  {\doibase 10.1103/PhysRevLett.125.074501} {\bibfield  {journal} {\bibinfo
  {journal} {Phys. Rev. Lett.}\ }\textbf {\bibinfo {volume} {125}},\ \bibinfo
  {pages} {074501} (\bibinfo {year} {2020})}\BibitemShut {NoStop}%
\bibitem [{\citenamefont {Yang}\ \emph
  {et~al.}(2020{\natexlab{a}})\citenamefont {Yang}, \citenamefont {Chong},
  \citenamefont {Wang}, \citenamefont {Verzicco}, \citenamefont {Shishkina},\
  and\ \citenamefont {Lohse}}]{yang20a}%
  \BibitemOpen
  \bibfield  {author} {\bibinfo {author} {\bibfnamefont {R.}~\bibnamefont
  {Yang}}, \bibinfo {author} {\bibfnamefont {K.-L.}\ \bibnamefont {Chong}},
  \bibinfo {author} {\bibfnamefont {Q.}~\bibnamefont {Wang}}, \bibinfo {author}
  {\bibfnamefont {R.}~\bibnamefont {Verzicco}}, \bibinfo {author}
  {\bibfnamefont {O.}~\bibnamefont {Shishkina}}, \ and\ \bibinfo {author}
  {\bibfnamefont {D.}~\bibnamefont {Lohse}},\ }\bibfield  {title} {\enquote
  {\bibinfo {title} {Periodically modulated thermal convection},}\ }\href
  {\doibase 10.1103/PhysRevLett.125.154502} {\bibfield  {journal} {\bibinfo
  {journal} {Phys. Rev. Lett.}\ }\textbf {\bibinfo {volume} {125}},\ \bibinfo
  {pages} {154502} (\bibinfo {year} {2020}{\natexlab{a}})}\BibitemShut
  {NoStop}%
\bibitem [{\citenamefont {Landau}\ and\ \citenamefont
  {Lifshitz}(2013)}]{landau13a}%
  \BibitemOpen
  \bibfield  {author} {\bibinfo {author} {\bibfnamefont {L.D.}\ \bibnamefont
  {Landau}}\ and\ \bibinfo {author} {\bibfnamefont {E.M.}\ \bibnamefont
  {Lifshitz}},\ }\href@noop {} {\emph {\bibinfo {title} {{Fluid Mechanics}}}},\
  \bibinfo {number} {v. 6}\ (\bibinfo  {publisher} {Elsevier Science, Oxford},\
  \bibinfo {year} {2013})\BibitemShut {NoStop}%
\bibitem [{\citenamefont {Gray}\ and\ \citenamefont
  {Giorgini}(1976)}]{gray76a}%
  \BibitemOpen
  \bibfield  {author} {\bibinfo {author} {\bibfnamefont {D.~D.}\ \bibnamefont
  {Gray}}\ and\ \bibinfo {author} {\bibfnamefont {A.}~\bibnamefont
  {Giorgini}},\ }\bibfield  {title} {\enquote {\bibinfo {title} {The validity
  of the {Boussinesq} approximation for liquids and gases},}\ }\href {\doibase
  10.1016/0017-9310(76)90168-X} {\bibfield  {journal} {\bibinfo  {journal}
  {Int. J. Heat Mass Transf.}\ }\textbf {\bibinfo {volume} {19}},\ \bibinfo
  {pages} {545} (\bibinfo {year} {1976})}\BibitemShut {NoStop}%
\bibitem [{\citenamefont {Zaitsev}\ and\ \citenamefont
  {Shliomis}(1971)}]{zaitsev71a}%
  \BibitemOpen
  \bibfield  {author} {\bibinfo {author} {\bibfnamefont {V.~M.}\ \bibnamefont
  {Zaitsev}}\ and\ \bibinfo {author} {\bibfnamefont {M.~I.}\ \bibnamefont
  {Shliomis}},\ }\bibfield  {title} {\enquote {\bibinfo {title} {Hydrodynamic
  fluctuations near convection threshold},}\ }\href
  {https://www.researchgate.net/profile/Mark-Shliomis/publication/253525252_Hydrodynamic_Fluctuations_near_the_Convection_Threshold/links/58e7edb9458515e30dcaf216/Hydrodynamic-Fluctuations-near-the-Convection-Threshold.pdf}
  {\bibfield  {journal} {\bibinfo  {journal} {Sov. Phys. JETP}\ }\textbf
  {\bibinfo {volume} {32}},\ \bibinfo {pages} {866} (\bibinfo {year}
  {1971})}\BibitemShut {NoStop}%
\bibitem [{\citenamefont {Graham}(1974)}]{graham74a}%
  \BibitemOpen
  \bibfield  {author} {\bibinfo {author} {\bibfnamefont {R.}~\bibnamefont
  {Graham}},\ }\bibfield  {title} {\enquote {\bibinfo {title} {Hydrodynamic
  fluctuations near the convection instability},}\ }\href {\doibase
  10.1103/PhysRevA.10.1762} {\bibfield  {journal} {\bibinfo  {journal} {Phys.
  Rev. A}\ }\textbf {\bibinfo {volume} {10}},\ \bibinfo {pages} {1762}
  (\bibinfo {year} {1974})}\BibitemShut {NoStop}%
\bibitem [{\citenamefont {Swift}\ and\ \citenamefont
  {Hohenberg}(1977)}]{swift77a}%
  \BibitemOpen
  \bibfield  {author} {\bibinfo {author} {\bibfnamefont {J.}~\bibnamefont
  {Swift}}\ and\ \bibinfo {author} {\bibfnamefont {P.~C.}\ \bibnamefont
  {Hohenberg}},\ }\bibfield  {title} {\enquote {\bibinfo {title} {Hydrodynamic
  fluctuations at the convective instability},}\ }\href {\doibase
  10.1103/PhysRevA.15.319} {\bibfield  {journal} {\bibinfo  {journal} {Phys.
  Rev. A}\ }\textbf {\bibinfo {volume} {15}},\ \bibinfo {pages} {319} (\bibinfo
  {year} {1977})}\BibitemShut {NoStop}%
\bibitem [{\citenamefont {Wu}\ \emph {et~al.}(1995)\citenamefont {Wu},
  \citenamefont {Ahlers},\ and\ \citenamefont {Cannell}}]{wu95a}%
  \BibitemOpen
  \bibfield  {author} {\bibinfo {author} {\bibfnamefont {M.}~\bibnamefont
  {Wu}}, \bibinfo {author} {\bibfnamefont {G.}~\bibnamefont {Ahlers}}, \ and\
  \bibinfo {author} {\bibfnamefont {D.~S.}\ \bibnamefont {Cannell}},\
  }\bibfield  {title} {\enquote {\bibinfo {title} {Thermally induced
  fluctuations below the onset of {Rayleigh-B{\'e}nard} convection},}\ }\href
  {\doibase 10.1103/PhysRevLett.75.1743} {\bibfield  {journal} {\bibinfo
  {journal} {Phys. Rev. Lett.}\ }\textbf {\bibinfo {volume} {75}},\ \bibinfo
  {pages} {1743} (\bibinfo {year} {1995})}\BibitemShut {NoStop}%
\bibitem [{\citenamefont {Pandey}\ \emph {et~al.}(2021)\citenamefont {Pandey},
  \citenamefont {Schumacher},\ and\ \citenamefont {Sreenivasan}}]{pandey21a}%
  \BibitemOpen
  \bibfield  {author} {\bibinfo {author} {\bibfnamefont {A.}~\bibnamefont
  {Pandey}}, \bibinfo {author} {\bibfnamefont {J.}~\bibnamefont {Schumacher}},
  \ and\ \bibinfo {author} {\bibfnamefont {K.~R.}\ \bibnamefont
  {Sreenivasan}},\ }\bibfield  {title} {\enquote {\bibinfo {title}
  {{Non-Boussinesq} low-{Prandtl}-number convection with a
  temperature-dependent thermal diffusivity},}\ }\href {\doibase
  10.3847/1538-4357/abd1d8} {\bibfield  {journal} {\bibinfo  {journal}
  {Astrophys. J.}\ }\textbf {\bibinfo {volume} {907}},\ \bibinfo {pages} {56}
  (\bibinfo {year} {2021})}\BibitemShut {NoStop}%
\bibitem [{\citenamefont {Bormann}(2001)}]{bormann01a}%
  \BibitemOpen
  \bibfield  {author} {\bibinfo {author} {\bibfnamefont {A.~S.}\ \bibnamefont
  {Bormann}},\ }\bibfield  {title} {\enquote {\bibinfo {title} {The onset of
  convection in the {Rayleigh-B{\'e}nard} problem for compressible fluids},}\
  }\href {\doibase 10.1007/s001610100039} {\bibfield  {journal} {\bibinfo
  {journal} {Contin. Mech. Thermodyn.}\ }\textbf {\bibinfo {volume} {13}},\
  \bibinfo {pages} {9} (\bibinfo {year} {2001})}\BibitemShut {NoStop}%
\bibitem [{\citenamefont {Risso}\ and\ \citenamefont
  {Cordero}(1993)}]{risso93a}%
  \BibitemOpen
  \bibfield  {author} {\bibinfo {author} {\bibfnamefont {D.}~\bibnamefont
  {Risso}}\ and\ \bibinfo {author} {\bibfnamefont {P.}~\bibnamefont
  {Cordero}},\ }\bibfield  {title} {\enquote {\bibinfo {title} {Empirical
  determination of the onset of convection for a hard disk system},}\ }in\
  \href {\doibase 10.1007/978-94-011-1906-1_20} {\emph {\bibinfo {booktitle}
  {Instabilities and Nonequilibrium Structures IV}}}\ (\bibinfo  {publisher}
  {Springer, Dordrecht},\ \bibinfo {year} {1993})\ p.\ \bibinfo {pages} {199}\BibitemShut
  {NoStop}%
\bibitem [{\citenamefont {Mareschal}\ and\ \citenamefont
  {Kestemont}(1987{\natexlab{a}})}]{mareschal87a}%
  \BibitemOpen
  \bibfield  {author} {\bibinfo {author} {\bibfnamefont {M.}~\bibnamefont
  {Mareschal}}\ and\ \bibinfo {author} {\bibfnamefont {E.}~\bibnamefont
  {Kestemont}},\ }\bibfield  {title} {\enquote {\bibinfo {title}
  {Experimental-evidence for convective rolls in finite two-dimensional
  molecular-models},}\ }\href {\doibase 10.1038/329427a0} {\bibfield  {journal}
  {\bibinfo  {journal} {Nature}\ }\textbf {\bibinfo {volume} {329}},\ \bibinfo
  {pages} {427} (\bibinfo {year} {1987}{\natexlab{a}})}\BibitemShut {NoStop}%
\bibitem [{\citenamefont {Mareschal}\ and\ \citenamefont
  {Kestemont}(1987{\natexlab{b}})}]{mareschal87b}%
  \BibitemOpen
  \bibfield  {author} {\bibinfo {author} {\bibfnamefont {M.}~\bibnamefont
  {Mareschal}}\ and\ \bibinfo {author} {\bibfnamefont {E.}~\bibnamefont
  {Kestemont}},\ }\bibfield  {title} {\enquote {\bibinfo {title} {Order and
  fluctuations in nonequilibrium molecular dynamics simulations of
  two-dimensional fluids},}\ }\href {\doibase 10.1007/BF01009540} {\bibfield
  {journal} {\bibinfo  {journal} {J. Stat. Phys.}\ }\textbf {\bibinfo {volume}
  {48}},\ \bibinfo {pages} {1187} (\bibinfo {year}
  {1987}{\natexlab{b}})}\BibitemShut {NoStop}%
\bibitem [{\citenamefont {Mareschal}\ \emph {et~al.}(1988)\citenamefont
  {Mareschal}, \citenamefont {Malek~Mansour}, \citenamefont {Puhl},\ and\
  \citenamefont {Kestemont}}]{mareschal88a}%
  \BibitemOpen
  \bibfield  {author} {\bibinfo {author} {\bibfnamefont {M.}~\bibnamefont
  {Mareschal}}, \bibinfo {author} {\bibfnamefont {M.}~\bibnamefont
  {Malek~Mansour}}, \bibinfo {author} {\bibfnamefont {A.}~\bibnamefont {Puhl}},
  \ and\ \bibinfo {author} {\bibfnamefont {E.}~\bibnamefont {Kestemont}},\
  }\bibfield  {title} {\enquote {\bibinfo {title} {Molecular dynamics versus
  hydrodynamics in a two-dimensional {Rayleigh-B\'enard} system},}\ }\href
  {\doibase 10.1103/PhysRevLett.61.2550} {\bibfield  {journal} {\bibinfo
  {journal} {Phys. Rev. Lett.}\ }\textbf {\bibinfo {volume} {61}},\ \bibinfo
  {pages} {2550} (\bibinfo {year} {1988})}\BibitemShut {NoStop}%
\bibitem [{\citenamefont {Rapaport}(1988)}]{rapaport88a}%
  \BibitemOpen
  \bibfield  {author} {\bibinfo {author} {\bibfnamefont {D.~C.}\ \bibnamefont
  {Rapaport}},\ }\bibfield  {title} {\enquote {\bibinfo {title}
  {Molecular-dynamics study of {Rayleigh-B\'enard} convection},}\ }\href
  {\doibase 10.1103/PhysRevLett.60.2480} {\bibfield  {journal} {\bibinfo
  {journal} {Phys. Rev. Lett.}\ }\textbf {\bibinfo {volume} {60}},\ \bibinfo
  {pages} {2480} (\bibinfo {year} {1988})}\BibitemShut {NoStop}%
\bibitem [{\citenamefont {Puhl}\ \emph {et~al.}(1989)\citenamefont {Puhl},
  \citenamefont {Mansour},\ and\ \citenamefont {Mareschal}}]{puhl89a}%
  \BibitemOpen
  \bibfield  {author} {\bibinfo {author} {\bibfnamefont {A.}~\bibnamefont
  {Puhl}}, \bibinfo {author} {\bibfnamefont {M.~Malek}\ \bibnamefont
  {Mansour}}, \ and\ \bibinfo {author} {\bibfnamefont {M.}~\bibnamefont
  {Mareschal}},\ }\bibfield  {title} {\enquote {\bibinfo {title} {Quantitative
  comparison of molecular dynamics with hydrodynamics in {Rayleigh-B\'enard}
  convection},}\ }\href {\doibase 10.1103/PhysRevA.40.1999} {\bibfield
  {journal} {\bibinfo  {journal} {Phys. Rev. A}\ }\textbf {\bibinfo {volume}
  {40}},\ \bibinfo {pages} {1999} (\bibinfo {year} {1989})}\BibitemShut
  {NoStop}%
\bibitem [{\citenamefont {Rapaport}(1992)}]{rapaport92a}%
  \BibitemOpen
  \bibfield  {author} {\bibinfo {author} {\bibfnamefont {D.~C.}\ \bibnamefont
  {Rapaport}},\ }\bibfield  {title} {\enquote {\bibinfo {title} {Unpredictable
  convection in a small box: Molecular-dynamics experiments},}\ }\href
  {\doibase 10.1103/PhysRevA.46.1971} {\bibfield  {journal} {\bibinfo
  {journal} {Phys. Rev. A}\ }\textbf {\bibinfo {volume} {46}},\ \bibinfo
  {pages} {1971} (\bibinfo {year} {1992})}\BibitemShut {NoStop}%
\bibitem [{\citenamefont {Rapaport}(2006)}]{rapaport06a}%
  \BibitemOpen
  \bibfield  {author} {\bibinfo {author} {\bibfnamefont {D.~C.}\ \bibnamefont
  {Rapaport}},\ }\bibfield  {title} {\enquote {\bibinfo {title} {Hexagonal
  convection patterns in atomistically simulated fluids},}\ }\href {\doibase
  https://doi.org/10.1103/PhysRevE.73.025301} {\bibfield  {journal} {\bibinfo
  {journal} {Phys. Rev. E}\ }\textbf {\bibinfo {volume} {73}},\ \bibinfo
  {pages} {025301} (\bibinfo {year} {2006})}\BibitemShut {NoStop}%
\bibitem [{\citenamefont {Yang}\ \emph
  {et~al.}(2020{\natexlab{b}})\citenamefont {Yang}, \citenamefont {Chen},
  \citenamefont {Verzicco},\ and\ \citenamefont {Lohse}}]{yang20b}%
  \BibitemOpen
  \bibfield  {author} {\bibinfo {author} {\bibfnamefont {Y.}~\bibnamefont
  {Yang}}, \bibinfo {author} {\bibfnamefont {W.}~\bibnamefont {Chen}}, \bibinfo
  {author} {\bibfnamefont {R.}~\bibnamefont {Verzicco}}, \ and\ \bibinfo
  {author} {\bibfnamefont {D.}~\bibnamefont {Lohse}},\ }\bibfield  {title}
  {\enquote {\bibinfo {title} {Multiple states and transport properties of
  double-diffusive convection turbulence},}\ }\href {\doibase
  10.1073/pnas.2005669117} {\bibfield  {journal} {\bibinfo  {journal} {Proc.
  Natl. Acad. Sci. USA}\ }\textbf {\bibinfo {volume} {117}},\ \bibinfo {pages}
  {14676} (\bibinfo {year} {2020}{\natexlab{b}})}\BibitemShut {NoStop}%
\bibitem [{\citenamefont {Avila}\ \emph {et~al.}(2011)\citenamefont {Avila},
  \citenamefont {Moxey}, \citenamefont {de~Lozar}, \citenamefont {Avila},
  \citenamefont {Barkley},\ and\ \citenamefont {Hof}}]{avila11a}%
  \BibitemOpen
  \bibfield  {author} {\bibinfo {author} {\bibfnamefont {K.}~\bibnamefont
  {Avila}}, \bibinfo {author} {\bibfnamefont {D.}~\bibnamefont {Moxey}},
  \bibinfo {author} {\bibfnamefont {A.}~\bibnamefont {de~Lozar}}, \bibinfo
  {author} {\bibfnamefont {M.}~\bibnamefont {Avila}}, \bibinfo {author}
  {\bibfnamefont {D.}~\bibnamefont {Barkley}}, \ and\ \bibinfo {author}
  {\bibfnamefont {B.}~\bibnamefont {Hof}},\ }\bibfield  {title} {\enquote
  {\bibinfo {title} {The onset of turbulence in pipe flow},}\ }\href {\doibase
  10.1126/science.1203223} {\bibfield  {journal} {\bibinfo  {journal}
  {Science}\ }\textbf {\bibinfo {volume} {333}},\ \bibinfo {pages} {192}
  (\bibinfo {year} {2011})}\BibitemShut {NoStop}%
\bibitem [{\citenamefont {Shi}\ \emph {et~al.}(2013)\citenamefont {Shi},
  \citenamefont {Avila},\ and\ \citenamefont {Hof}}]{shi13a}%
  \BibitemOpen
  \bibfield  {author} {\bibinfo {author} {\bibfnamefont {L.}~\bibnamefont
  {Shi}}, \bibinfo {author} {\bibfnamefont {M.}~\bibnamefont {Avila}}, \ and\
  \bibinfo {author} {\bibfnamefont {B.}~\bibnamefont {Hof}},\ }\bibfield
  {title} {\enquote {\bibinfo {title} {Scale invariance at the onset of
  turbulence in {Couette} flow},}\ }\href {\doibase
  10.1103/PhysRevLett.110.204502} {\bibfield  {journal} {\bibinfo  {journal}
  {Phys. Rev. Lett.}\ }\textbf {\bibinfo {volume} {110}},\ \bibinfo {pages}
  {204502} (\bibinfo {year} {2013})}\BibitemShut {NoStop}%
\bibitem [{\citenamefont {Lemoult}\ \emph {et~al.}(2016)\citenamefont
  {Lemoult}, \citenamefont {Shi}, \citenamefont {Avila}, \citenamefont
  {Jalikop}, \citenamefont {Avila},\ and\ \citenamefont {Hof}}]{lemoult16a}%
  \BibitemOpen
  \bibfield  {author} {\bibinfo {author} {\bibfnamefont {G.}~\bibnamefont
  {Lemoult}}, \bibinfo {author} {\bibfnamefont {L.}~\bibnamefont {Shi}},
  \bibinfo {author} {\bibfnamefont {K.}~\bibnamefont {Avila}}, \bibinfo
  {author} {\bibfnamefont {S.~V.}\ \bibnamefont {Jalikop}}, \bibinfo {author}
  {\bibfnamefont {M.}~\bibnamefont {Avila}}, \ and\ \bibinfo {author}
  {\bibfnamefont {B.}~\bibnamefont {Hof}},\ }\bibfield  {title} {\enquote
  {\bibinfo {title} {Directed percolation phase transition to sustained
  turbulence in {Couette} flow},}\ }\href {\doibase 10.1038/nphys3675}
  {\bibfield  {journal} {\bibinfo  {journal} {Nature Physics}\ }\textbf
  {\bibinfo {volume} {12}},\ \bibinfo {pages} {254} (\bibinfo {year}
  {2016})}\BibitemShut {NoStop}%
\bibitem [{\citenamefont {Klotz}\ \emph {et~al.}(2022)\citenamefont {Klotz},
  \citenamefont {Lemoult}, \citenamefont {Avila},\ and\ \citenamefont
  {Hof}}]{klotz22a}%
  \BibitemOpen
  \bibfield  {author} {\bibinfo {author} {\bibfnamefont {L.}~\bibnamefont
  {Klotz}}, \bibinfo {author} {\bibfnamefont {G.}~\bibnamefont {Lemoult}},
  \bibinfo {author} {\bibfnamefont {K.}~\bibnamefont {Avila}}, \ and\ \bibinfo
  {author} {\bibfnamefont {B.}~\bibnamefont {Hof}},\ }\bibfield  {title}
  {\enquote {\bibinfo {title} {Phase transition to turbulence in spatially
  extended shear flows},}\ }\href {\doibase 10.1103/PhysRevLett.128.014502}
  {\bibfield  {journal} {\bibinfo  {journal} {Phys. Rev. Lett.}\ }\textbf
  {\bibinfo {volume} {128}},\ \bibinfo {pages} {014502} (\bibinfo {year}
  {2022})}\BibitemShut {NoStop}%
\bibitem [{\citenamefont {Zhang}\ and\ \citenamefont {Fan}(2009)}]{zhang09a}%
  \BibitemOpen
  \bibfield  {author} {\bibinfo {author} {\bibfnamefont {J.}~\bibnamefont
  {Zhang}}\ and\ \bibinfo {author} {\bibfnamefont {J.}~\bibnamefont {Fan}},\
  }\bibfield  {title} {\enquote {\bibinfo {title} {{Monte Carlo} simulation of
  thermal fluctuations below the onset of {Rayleigh-B\'enard} convection},}\
  }\href {\doibase 10.1103/PhysRevE.79.056302} {\bibfield  {journal} {\bibinfo
  {journal} {Phys. Rev. E}\ }\textbf {\bibinfo {volume} {79}},\ \bibinfo
  {pages} {056302} (\bibinfo {year} {2009})}\BibitemShut {NoStop}%
\bibitem [{\citenamefont {Zhang}\ and\ \citenamefont
  {\"Onskog}(2017)}]{zhang17b}%
  \BibitemOpen
  \bibfield  {author} {\bibinfo {author} {\bibfnamefont {J.}~\bibnamefont
  {Zhang}}\ and\ \bibinfo {author} {\bibfnamefont {T.}~\bibnamefont
  {\"Onskog}},\ }\bibfield  {title} {\enquote {\bibinfo {title} {Langevin
  equation elucidates the mechanism of the {Rayleigh-B\'enard} instability by
  coupling molecular motions and macroscopic fluctuations},}\ }\href {\doibase
  10.1103/PhysRevE.96.043104} {\bibfield  {journal} {\bibinfo  {journal} {Phys.
  Rev. E}\ }\textbf {\bibinfo {volume} {96}},\ \bibinfo {pages} {043104}
  (\bibinfo {year} {2017})}\BibitemShut {NoStop}%
\bibitem [{\citenamefont {Mulero}(2008)}]{mulero08a}%
  \BibitemOpen
  \bibfield  {author} {\bibinfo {author} {\bibfnamefont {A.}~\bibnamefont
  {Mulero}},\ }\href
  {https://link.springer.com/book/10.1007%2F978-3-540-78767-9} {\emph {\bibinfo
  {title} {Theory and {Simulation} of {Hard-Sphere} {Fluids} and {Related}
  Systems}}},\ \bibinfo {series} {Lecture Notes in Physics}, Vol.\ \bibinfo
  {volume} {753}\ (\bibinfo  {publisher} {SpringerVerlag},\ \bibinfo {year}
  {2008})\BibitemShut {NoStop}%
\bibitem [{\citenamefont {Sz{\'a}sz}(2000)}]{szasz00a}%
  \BibitemOpen
  \bibinfo {editor} {\bibfnamefont {D.}~\bibnamefont {Sz{\'a}sz}},\ ed.,\ \href
  {https://www.springer.com/gp/book/9783540676201} {\emph {\bibinfo {title}
  {Hard Ball Systems and the Lorentz Gas}}}\ (\bibinfo  {publisher}
  {Springer-Verlag},\ \bibinfo {address} {Berlin},\ \bibinfo {year}
  {2000})\BibitemShut {NoStop}%
\bibitem [{\citenamefont {Hurtado}\ and\ \citenamefont
  {Garrido}(2020)}]{hurtado20a}%
  \BibitemOpen
  \bibfield  {author} {\bibinfo {author} {\bibfnamefont {P.~I.}\ \bibnamefont
  {Hurtado}}\ and\ \bibinfo {author} {\bibfnamefont {P.~L.}\ \bibnamefont
  {Garrido}},\ }\bibfield  {title} {\enquote {\bibinfo {title} {Simulations of
  transport in hard particle systems},}\ }\href {\doibase
  10.1007/s10955-019-02469-z} {\bibfield  {journal} {\bibinfo  {journal} {J.
  Stat. Phys.}\ }\textbf {\bibinfo {volume} {180}},\ \bibinfo {pages} {474}
  (\bibinfo {year} {2020})}\BibitemShut {NoStop}%
\bibitem [{\citenamefont {Livi}\ and\ \citenamefont {Politi}(2017)}]{livi17a}%
  \BibitemOpen
  \bibfield  {author} {\bibinfo {author} {\bibfnamefont {R.}~\bibnamefont
  {Livi}}\ and\ \bibinfo {author} {\bibfnamefont {P.}~\bibnamefont {Politi}},\
  }\href {\doibase 10.1017/9781107278974} {\emph {\bibinfo {title}
  {Nonequilibrium Statistical Physics: A Modern Perspective}}}\ (\bibinfo
  {publisher} {Cambridge University Press, Cambridge},\ \bibinfo {year}
  {2017})\BibitemShut {NoStop}%
\bibitem [{\citenamefont {Dhar}(2008)}]{dhar08a}%
  \BibitemOpen
  \bibfield  {author} {\bibinfo {author} {\bibfnamefont {A.}~\bibnamefont
  {Dhar}},\ }\bibfield  {title} {\enquote {\bibinfo {title} {Heat transport in
  low-dimensional systems},}\ }\href
  {http://www.tandfonline.com/doi/abs/10.1080/00018730802538522} {\bibfield
  {journal} {\bibinfo  {journal} {Adv. Phys.}\ }\textbf {\bibinfo {volume}
  {57}},\ \bibinfo {pages} {457} (\bibinfo {year} {2008})}\BibitemShut
  {NoStop}%
\bibitem [{\citenamefont {Lepri}\ \emph {et~al.}(2003)\citenamefont {Lepri},
  \citenamefont {Livi},\ and\ \citenamefont {Politi}}]{lepri03a}%
  \BibitemOpen
  \bibfield  {author} {\bibinfo {author} {\bibfnamefont {S.}~\bibnamefont
  {Lepri}}, \bibinfo {author} {\bibfnamefont {R.}~\bibnamefont {Livi}}, \ and\
  \bibinfo {author} {\bibfnamefont {A.}~\bibnamefont {Politi}},\ }\bibfield
  {title} {\enquote {\bibinfo {title} {Thermal conduction in classical
  low-dimensional lattices},}\ }\href {\doibase 10.1016/S0370-1573(02)00558-6}
  {\bibfield  {journal} {\bibinfo  {journal} {Phys. Rep.}\ }\textbf {\bibinfo
  {volume} {377}},\ \bibinfo {pages} {1--80} (\bibinfo {year}
  {2003})}\BibitemShut {NoStop}%
\bibitem [{\citenamefont {Bonetto}\ \emph {et~al.}(2000)\citenamefont
  {Bonetto}, \citenamefont {Lebowitz},\ and\ \citenamefont
  {Rey-Bellet}}]{bonetto00a}%
  \BibitemOpen
  \bibfield  {author} {\bibinfo {author} {\bibfnamefont {F.}~\bibnamefont
  {Bonetto}}, \bibinfo {author} {\bibfnamefont {J.~L.}\ \bibnamefont
  {Lebowitz}}, \ and\ \bibinfo {author} {\bibfnamefont {L.}~\bibnamefont
  {Rey-Bellet}},\ }\enquote {\bibinfo {title} {Mathematical physics 2000},}\ \
  (\bibinfo  {publisher} {Imperial College Press},\ \bibinfo {address}
  {London},\ \bibinfo {year} {2000})\ Chap.\ \bibinfo {chapter} {{F}ourier's
  law: {A} challenge for theorists}, p.\ \bibinfo {pages} {128}\BibitemShut
  {NoStop}%
\bibitem [{\citenamefont {del Pozo}\ \emph
  {et~al.}(2015{\natexlab{a}})\citenamefont {del Pozo}, \citenamefont
  {Garrido},\ and\ \citenamefont {Hurtado}}]{pozo15b}%
  \BibitemOpen
  \bibfield  {author} {\bibinfo {author} {\bibfnamefont {J.~J.}\ \bibnamefont
  {del Pozo}}, \bibinfo {author} {\bibfnamefont {P.~L.}\ \bibnamefont
  {Garrido}}, \ and\ \bibinfo {author} {\bibfnamefont {P.~I.}\ \bibnamefont
  {Hurtado}},\ }\bibfield  {title} {\enquote {\bibinfo {title} {Scaling laws
  and bulk-boundary decoupling in heat flow},}\ }\href
  {http://journals.aps.org/pre/abstract/10.1103/PhysRevE.91.032116} {\bibfield
  {journal} {\bibinfo  {journal} {Phys. Rev. E}\ }\textbf {\bibinfo {volume}
  {91}},\ \bibinfo {pages} {032116} (\bibinfo {year}
  {2015}{\natexlab{a}})}\BibitemShut {NoStop}%
\bibitem [{\citenamefont {del Pozo}\ \emph
  {et~al.}(2015{\natexlab{b}})\citenamefont {del Pozo}, \citenamefont
  {Garrido},\ and\ \citenamefont {Hurtado}}]{pozo15a}%
  \BibitemOpen
  \bibfield  {author} {\bibinfo {author} {\bibfnamefont {J.~J.}\ \bibnamefont
  {del Pozo}}, \bibinfo {author} {\bibfnamefont {P.~L.}\ \bibnamefont
  {Garrido}}, \ and\ \bibinfo {author} {\bibfnamefont {P.~I.}\ \bibnamefont
  {Hurtado}},\ }\bibfield  {title} {\enquote {\bibinfo {title} {Probing local
  equilibrium in nonequilibrium fluids},}\ }\href
  {http://journals.aps.org/pre/abstract/10.1103/PhysRevE.92.022117} {\bibfield
  {journal} {\bibinfo  {journal} {Phys. Rev. E}\ }\textbf {\bibinfo {volume}
  {92}},\ \bibinfo {pages} {022117} (\bibinfo {year}
  {2015}{\natexlab{b}})}\BibitemShut {NoStop}%
\bibitem [{\citenamefont {Hurtado}\ and\ \citenamefont
  {Garrido}(2016)}]{hurtado16a}%
  \BibitemOpen
  \bibfield  {author} {\bibinfo {author} {\bibfnamefont {P.~I.}\ \bibnamefont
  {Hurtado}}\ and\ \bibinfo {author} {\bibfnamefont {P.~L.}\ \bibnamefont
  {Garrido}},\ }\bibfield  {title} {\enquote {\bibinfo {title} {A violation of
  universality in anomalous {Fourier's} law},}\ }\href {\doibase
  10.1038/srep38823} {\bibfield  {journal} {\bibinfo  {journal} {Scientific
  Reports}\ }\textbf {\bibinfo {volume} {6}},\ \bibinfo {pages} {38823}
  (\bibinfo {year} {2016})}\BibitemShut {NoStop}%
\bibitem [{\citenamefont {Henderson}(1977)}]{henderson77a}%
  \BibitemOpen
  \bibfield  {author} {\bibinfo {author} {\bibfnamefont {D.}~\bibnamefont
  {Henderson}},\ }\bibfield  {title} {\enquote {\bibinfo {title} {{Monte-Carlo}
  and perturbation-theory studies of equation of state of 2-dimensional
  {Lennard-Jones} fluid},}\ }\href {\doibase 10.1080/00268977700101741}
  {\bibfield  {journal} {\bibinfo  {journal} {Mol. Phys.}\ }\textbf {\bibinfo
  {volume} {34}},\ \bibinfo {pages} {301} (\bibinfo {year} {1977})}\BibitemShut
  {NoStop}%
\bibitem [{Note1()}]{Note1}%
  \BibitemOpen
  \bibinfo {note} {Previous works \cite {wesfreid78a} have used the maximum of
  the velocity field as order parameter. This observable is subject to stronger
  fluctuations, while the hydrodynamic kinetic energy has better averaging
  properties and allows for a more accurate characterization of the RB
  transition.}\BibitemShut {Stop}%
\bibitem [{Note2()}]{Note2}%
  \BibitemOpen
  \bibinfo {note} {Note that the observation that $\la J_g\ra < \la J_0\ra$ for $1<T<T_J(g)$ implies an estimated Nusselt number $\text{Nu}\approx \la J_g\ra /\la J_0\ra<1$, see Appendix \ref{appB2}, in apparent contradiction with a bound $\text{Nu}>1$ obtained within the Oberbeck-Boussinesq hydrodynamics approximation. However, as shown in Figs.~\ref{figT} and \ref{figrho} and the associated discussion, our fluid model is clearly compressible and non-Oberbeck-Boussinesq in the parameter range studied, and hence there is no contradiction with the aforementioned bound. This is another instance of the ill-defined behavior of the dimensionless numbers of linear hydrodynamics under strong-driving conditions.}\BibitemShut
  {NoStop}%
\bibitem [{\citenamefont {{Mertens}}\ and\ \citenamefont
  {{Moore}}(2012)}]{mertens12a}%
  \BibitemOpen
  \bibfield  {author} {\bibinfo {author} {\bibfnamefont {S.}~\bibnamefont
  {{Mertens}}}\ and\ \bibinfo {author} {\bibfnamefont {C.}~\bibnamefont
  {{Moore}}},\ }\bibfield  {title} {\enquote {\bibinfo {title} {{Continuum
  percolation thresholds in two dimensions}},}\ }\href {\doibase
  10.1103/PhysRevE.86.061109} {\bibfield  {journal} {\bibinfo  {journal} {Phys.
  Rev. E}\ }\textbf {\bibinfo {volume} {86}},\ \bibinfo {pages} {061109}
  (\bibinfo {year} {2012})}\BibitemShut {NoStop}%
\bibitem [{\citenamefont {Stalling}\ and\ \citenamefont
  {Hege}(1995)}]{stalling95a}%
  \BibitemOpen
  \bibfield  {author} {\bibinfo {author} {\bibfnamefont {D.}~\bibnamefont
  {Stalling}}\ and\ \bibinfo {author} {\bibfnamefont {H.~C.}\ \bibnamefont
  {Hege}},\ }\bibfield  {title} {\enquote {\bibinfo {title} {Fast and
  resolution independent line integral convolution},}\ }in\ \href
  {https://dl.acm.org/citation.cfm?id=218380&picked=prox} {\emph {\bibinfo
  {booktitle} {Proceedings of the 22nd annual conference on computer graphics
  and interactive techniques (SIGGRAPH'95)}}},\ \bibinfo {editor} {edited by\
  \bibinfo {editor} {\bibfnamefont {S.~G.}\ \bibnamefont {Mair}}\ and\ \bibinfo
  {editor} {\bibfnamefont {R.}~\bibnamefont {Cook}}}\ (\bibinfo  {publisher}
  {ACM},\ \bibinfo {address} {New York},\ \bibinfo {year} {1995})\ p.\ \bibinfo
  {pages} {249}\BibitemShut {NoStop}%
\bibitem [{\citenamefont {Cabral}\ and\ \citenamefont
  {Leedom}(1993)}]{cabral93a}%
  \BibitemOpen
  \bibfield  {author} {\bibinfo {author} {\bibfnamefont {B.}~\bibnamefont
  {Cabral}}\ and\ \bibinfo {author} {\bibfnamefont {L.~C.}\ \bibnamefont
  {Leedom}},\ }\bibfield  {title} {\enquote {\bibinfo {title} {Imaging vector
  fields using line integral convolution},}\ }\href
  {http://cs.brown.edu/courses/csci2370/2000/1999/cabral.pdf} {\bibfield
  {journal} {\bibinfo  {journal} {Computer Graphics}\ }\textbf {\bibinfo
  {volume} {27}},\ \bibinfo {pages} {263} (\bibinfo {year} {1993})}\BibitemShut
  {NoStop}%
\bibitem [{\citenamefont {Ortiz~de Zarate}\ and\ \citenamefont
  {Sengers}(2006)}]{zarate06a}%
  \BibitemOpen
  \bibfield  {author} {\bibinfo {author} {\bibfnamefont {J.M.}\ \bibnamefont
  {Ortiz~de Zarate}}\ and\ \bibinfo {author} {\bibfnamefont {J.V.}\
  \bibnamefont {Sengers}},\ }\href@noop {} {\emph {\bibinfo {title}
  {Hydrodynamic fluctuations in fluids and fluid mixtures}}}\ (\bibinfo
  {publisher} {Elsevier, Amsterdam},\ \bibinfo {year} {2006})\BibitemShut
  {NoStop}%
\bibitem [{\citenamefont {Berge}\ and\ \citenamefont
  {Dubois}(1974)}]{berge74a}%
  \BibitemOpen
  \bibfield  {author} {\bibinfo {author} {\bibfnamefont {P.}~\bibnamefont
  {Berge}}\ and\ \bibinfo {author} {\bibfnamefont {M.}~\bibnamefont {Dubois}},\
  }\bibfield  {title} {\enquote {\bibinfo {title} {Convective velocity field in
  the {Rayleigh-B{\'e}nard} instability: experimental results},}\ }\href
  {\doibase 10.1103/PhysRevLett.32.1041} {\bibfield  {journal} {\bibinfo
  {journal} {Phys. Rev. Lett.}\ }\textbf {\bibinfo {volume} {32}},\ \bibinfo
  {pages} {1041} (\bibinfo {year} {1974})}\BibitemShut {NoStop}%
\bibitem [{\citenamefont {Wesfreid}\ \emph {et~al.}(1978)\citenamefont
  {Wesfreid}, \citenamefont {Pomeau}, \citenamefont {Dubois}, \citenamefont
  {Normand},\ and\ \citenamefont {Berg{\'e}}}]{wesfreid78a}%
  \BibitemOpen
  \bibfield  {author} {\bibinfo {author} {\bibfnamefont {J.}~\bibnamefont
  {Wesfreid}}, \bibinfo {author} {\bibfnamefont {Y.}~\bibnamefont {Pomeau}},
  \bibinfo {author} {\bibfnamefont {M.}~\bibnamefont {Dubois}}, \bibinfo
  {author} {\bibfnamefont {C.}~\bibnamefont {Normand}}, \ and\ \bibinfo
  {author} {\bibfnamefont {P.}~\bibnamefont {Berg{\'e}}},\ }\bibfield  {title}
  {\enquote {\bibinfo {title} {Critical effects in {Rayleigh-B{\'e}nard}
  convection},}\ }\href {\doibase 10.1051/jphys:01978003907072500} {\bibfield
  {journal} {\bibinfo  {journal} {J. Phys. France}\ }\textbf {\bibinfo {volume}
  {39}},\ \bibinfo {pages} {725} (\bibinfo {year} {1978})}\BibitemShut
  {NoStop}%
\bibitem [{\citenamefont {Behringer}\ and\ \citenamefont
  {Ahlers}(1977)}]{behringer77a}%
  \BibitemOpen
  \bibfield  {author} {\bibinfo {author} {\bibfnamefont {R.~P.}\ \bibnamefont
  {Behringer}}\ and\ \bibinfo {author} {\bibfnamefont {G.}~\bibnamefont
  {Ahlers}},\ }\bibfield  {title} {\enquote {\bibinfo {title} {Heat transport
  and critical slowing down near the {Rayleigh-B{\'e}nard} instability in
  cylindrical containers},}\ }\href {\doibase 10.1016/0375-9601(77)90431-5}
  {\bibfield  {journal} {\bibinfo  {journal} {Phys. Lett. A}\ }\textbf
  {\bibinfo {volume} {62}},\ \bibinfo {pages} {329} (\bibinfo {year}
  {1977})}\BibitemShut {NoStop}%
\bibitem [{\citenamefont {Cordero}\ and\ \citenamefont
  {Risso}(1997)}]{cordero97a}%
  \BibitemOpen
  \bibfield  {author} {\bibinfo {author} {\bibfnamefont {P.}~\bibnamefont
  {Cordero}}\ and\ \bibinfo {author} {\bibfnamefont {D.}~\bibnamefont
  {Risso}},\ }\bibfield  {title} {\enquote {\bibinfo {title} {Microscopic
  computer simulation of fluids},}\ }in\ \href {\doibase
  10.1007/978-3-662-14148-9} {\emph {\bibinfo {booktitle} {{Fourth Granada
  Lectures on Computational Physics}}}},\ \bibinfo {series} {{Lecture Notes in
  Physics}}, Vol.\ \bibinfo {volume} {493},\ \bibinfo {editor} {edited by\
  \bibinfo {editor} {\bibfnamefont {P.L.}\ \bibnamefont {Garrido}}\ and\
  \bibinfo {editor} {\bibfnamefont {J.}~\bibnamefont {Marro}}}\ (\bibinfo
  {publisher} {Springer Berlin, Heidelberg},\ \bibinfo {year} {1997})\BibitemShut {NoStop}%
\bibitem [{\citenamefont {Press}\ \emph {et~al.}(1996)\citenamefont {Press},
  \citenamefont {Teukolsky}, \citenamefont {Vetterling},\ and\ \citenamefont
  {Flannery}}]{press96a}%
  \BibitemOpen
  \bibfield  {author} {\bibinfo {author} {\bibfnamefont {W.~H.}\ \bibnamefont
  {Press}}, \bibinfo {author} {\bibfnamefont {S.~A.}\ \bibnamefont
  {Teukolsky}}, \bibinfo {author} {\bibfnamefont {W.~T.}\ \bibnamefont
  {Vetterling}}, \ and\ \bibinfo {author} {\bibfnamefont {B.~P.}\ \bibnamefont
  {Flannery}},\ }\href {http://www.nrbook.com/a/bookf90pdf.html} {\emph
  {\bibinfo {title} {Numerical recipes in {Fortran} 90}}},\ Vol.~\bibinfo
  {volume} {2}\ (\bibinfo  {publisher} {Cambridge University Press},\ \bibinfo
  {address} {Cambridge},\ \bibinfo {year} {1996})\BibitemShut {NoStop}%
\end{thebibliography}
\end{document}